\documentclass[final,authoryear,5p]{elsarticle}

 \usepackage{graphicx}

\usepackage{amssymb}
\usepackage{marvosym}
\usepackage{pbox}

\journal{Advances in Space Research}

\begin{document}

\begin{frontmatter}


\title{The Castalia Mission to Main Belt Comet 133P/Elst-Pizarro}



\author[cs]{C. Snodgrass\corref{cor}}
\cortext[cor]{Corresponding author}
\ead{colin.snodgrass@open.ac.uk}

\author[gj]{G. H. Jones}
\author[hb]{H. Boehnhardt}
\author[ag]{A. Gibbings}
\author[ag]{M. Homeister}
\author[na]{N. Andre}
\author[pb]{P. Beck}
\author[msb]{M. S. Bentley}
\author[ib]{I.~Bertini}
\author[nb]{N. Bowles}
\author[mtc]{M. T. Capria}
\author[cc]{C. Carr}
\author[mc]{M. Ceriotti}
\author[gj]{A. J. Coates}
\author[mtc,ar]{V. Della Corte}
\author[nb]{K.~L.~Donaldson~Hanna}
\author[af]{A. Fitzsimmons}
\author[pg]{P. J. Guti{\'e}rrez}
\author[oh]{O.R. Hainaut}
\author[pb]{A. Herique}
\author[hb]{M. Hilchenbach}
\author[hhh,hhh1]{H.~H.~Hsieh}
\author[ej]{E. Jehin}
\author[ok]{O.~Karatekin}
\author[pb,wk]{W. Kofman}
\author[pg]{L. M. Lara}
\author[ag]{K. Laudan}
\author[jl]{J. Licandro}
\author[sl]{S. C. Lowry}
\author[fm]{F.~Marzari}
\author[cc]{A. Masters}
\author[km]{K.~J.~Meech}
\author[pg]{F. Moreno}
\author[cs]{A. Morse}
\author[ro]{R. Orosei}
\author[ap]{A. Pack}
\author[dp]{D. Plettemeier}
\author[dina]{D.~Prialnik}
\author[mtc,ar]{A.~Rotundi}
\author[ka]{M.~Rubin}
\author[jps]{J. P. S{\'a}nchez}
\author[cs]{S. Sheridan}
\author[mt]{M. Trieloff}
\author[ag]{A. Winterboer}

\address[cs]{Planetary and Space Sciences, School of Physical Sciences, The Open University, Milton Keynes, MK7 6AA, UK}
\address[gj]{Mullard Space Science Laboratory, University College London, Holmbury St Mary, Dorking, Surrey, RH5 6NT, UK}
\address[hb]{Max-Planck-Institut f{\"u}r Sonnensystemforschung, Justus-von-Liebig-Weg 3, 37077 G{\"o}ttingen, Germany}
\address[ag]{OHB System AG, Universit{\"a}tsallee 27-29, D-28359 Bremen, Germany}
\address[na]{Institut de Recherche en Astrophysique et Plan{\'e}tologie, Centre National de la Recherche Scientifique, Universit{\'e} Paul Sabatier Toulouse, 9 avenue du colonel Roche, BP 44346 31028 Cedex 4 Toulouse, France}
\address[pb]{Univ. Grenoble Alpes, CNRS, IPAG, F-38000 Grenoble, France}
\address[msb]{ Space Research Institute, Austrian Academy of Sciences, Schmiedlstrasse 6, 8042 Graz, Austria }
\address[ib]{Department of Physics and Astronomy `Galileo Galilei', University of Padova, Vic. Osservatorio 3, 35122 Padova, Italy}
\address[nb]{Department of Physics, Clarendon Laboratory, University of Oxford, Parks Road, Oxford, OX1 3PU, UK}
\address[mtc]{Istituto di Astrofisica e Planetologia Spaziali-INAF, Via Fosso del Cavaliere 100, 00133 Roma, Italy}
\address[cc]{The Blackett Laboratory, Imperial College London, Prince Consort Road, London, SW7 2AZ, UK}
\address[mc]{School of Engineering, University of Glasgow, Glasgow, G12 8QQ, UK}
\address[ar]{Dip. Scienze e Tecnologie, Universit{\`a} degli Studi di Napoli Parthenope, CDN, IC4, 80143 - Napoli, Italy}
\address[af]{Astrophysics Research Centre, School of Mathematics and Physics, Queen's University Belfast, Belfast, BT7 1NN, UK}
\address[pg]{Instituto de Astrof\'{i}sica de Andaluc\'{i}a, CSIC, Glorieta de la Astronom\'{i}a s/n, 18008 Granada, Spain}
\address[oh]{European Southern Observatory, Karl-Schwarzschild-Strasse 2, 85748 Garching-bei-M\"unchen, Germany}
\address[hhh]{Planetary Science Institute, 1700 East Fort Lowell Rd., Suite 106, Tucson, Arizona 86719, USA}
\address[hhh1]{Academia Sinica Institute of Astronomy and Astrophysics, P.O. Box 23-141, Taipei 10617, Taiwan}
\address[ej]{Institut d'Astrophysique et de G\'{e}ophysique, Universit\'{e} de Li\`{e}ge, Sart-Tilman, B-4000, Li\`{e}ge, Belgium}
\address[ok]{Royal Observatory of Belgium, Av. Circulaire 3, B-1180 Brussels, Belgium}
\address[wk]{Space Research Centre of the Polish Academy of Sciences, F-00-716 Warsaw, Poland}
\address[jl]{ Instituto de Astrof\'{\i}sica de Canarias (IAC), C\/V\'{\i}a L\'{a}ctea s\/n, 38205 La Laguna, Spain}
\address[sl]{Centre for Astrophysics and Planetary Science, School of Physical Sciences, The University of Kent, Canterbury, CT2 7NH, UK}
\address[fm]{Department of Physics, University of Padova, I-35131 Padova, Italy}
\address[km]{Institute for Astronomy, 2680 Woodlawn Drive, Honolulu, HI, 96822, USA}
\address[ro]{Istituto di Radioastronomia, Istituto Nazionale di Astrofisica,  Via Piero Gobetti 101, 40129 Bologna, Italy}
\address[ap]{Georg-August-Universit{\"a}t, Geowissenschaftliches Zentrum, Goldschmidtstrasse 1, 37077 G{\"o}ttingen, Germany}
\address[dp]{Technische Universit{\"a}t Dresden, Helmholtzstr. 10, 01069, Dresden, Germany}
\address[dina]{Department of Geosciences, Tel Aviv University, Ramat Aviv, Tel Aviv 69978, Israel}
\address[ka]{Physikalisches Institut, University of Bern, Sidlerstrasse 5, 3012 Bern, Switzerland}
\address[jps]{Space Research Group, Centre of Autonomous and Cyber-Physical Systems, Cranfield University, MK43 0AL, UK}
\address[mt]{Klaus-Tschira-Labor f{\"u}r Kosmochemie, Institut f{\"u}r Geowissenschaften, Universit{\"a}t  Heidelberg, Im Neuenheimer Feld 234-236, 69120 Heidelberg, Germany}

\begin{abstract}
We describe {\it Castalia}, a proposed mission to rendezvous with a Main Belt Comet (MBC), 133P/Elst-Pizarro. MBCs are a recently discovered population of apparently icy bodies within the main asteroid belt between Mars and Jupiter, which may represent the remnants of the population which supplied the early Earth with water. {\it Castalia} will perform the first exploration of this population by characterising 133P in detail, solving the puzzle of the MBC's activity, and making the first in situ measurements of water in the asteroid belt. In many ways a successor to ESA's highly successful {\it Rosetta} mission, {\it Castalia} will allow direct comparison between very different classes of comet, including measuring critical isotope ratios, plasma and dust properties. It will also feature the first radar system to visit a minor body, mapping the ice in the interior. {\it Castalia} was proposed, in slightly different versions, to the ESA M4 and M5 calls within the Cosmic Vision programme. We describe the science motivation for the mission, the measurements required to achieve the scientific goals, and the proposed instrument payload and spacecraft to achieve these.
\end{abstract}


\end{frontmatter}

\parindent=0.5 cm

\section{Introduction}

Our knowledge of the planets and small bodies of our Solar System increases via four routes: The steady flow of information from telescopic observations, laboratory analyses of accessible extraterrestrial materials, computer simulations, and the occasional great leaps forward that result from spacecraft exploration. The minor bodies, comets and asteroids, hold important clues to understand the process of planet formation, and a number of missions have visited them in recent years, with results reported throughout this special issue. A particular success has been the European Space Agency (ESA)'s {\it Rosetta} mission to comet 67P/Churyumov-Gerasimenko, which, among many highlights, demonstrated that this comet is a very primitive body, but its water does not match Earth's \citep{Altwegg-DH} -- a surprising result that indicates that Jupiter Family comets cannot be the main source of Earth's water as previously thought.

In the meantime, in the years since {\it Rosetta} was selected, built and launched, telescopic surveys have discovered a number of puzzling minor bodies that cross the boundary between asteroids and comets. The so-called Main Belt Comets (MBCs) have asteroid-like orbits but comet-like appearance and behaviour, and represent a new population that may better sample the source region of Earth's water. They therefore represent a natural  destination for the next comet mission.

The medium class of ESA missions represent the best opportunity to continue European leadership in the exploration of the small bodies of our Solar System; with currently planned larger missions for the next decades expected to visit Jupiter and provide the next generation of space observatory, there will not be another {\it Rosetta}-sized mission in the foreseeable future. Proposed to the M4 and M5 calls of the ESA Cosmic Vision programme, {\it Castalia} is a medium-sized mission that continues {\it Rosetta}'s legacy by rendezvousing with a MBC. Necessarily simpler than {\it Rosetta}, the mission has no lander and a reduced instrument complement, and makes use of advances in spacecraft autonomy to greatly simplify operations. This paper describes the scientific motivation for visiting a MBC (section \ref{sec:sci}), the necessary measurements (section \ref{sec:req}), and the proposed target and mission  (sections  \ref{sec:target} and \ref{sec:mission}), instruments (section \ref{sec:ins}),  and spacecraft (section \ref{sec:spacecraft}) to achieve this, based on the M5 version of the mission proposal (with improvements compared to the M4 version described as appropriate). Alternative approaches to a MBC mission were also proposed to M3 ({\it Caroline} -- \citealt{Caroline}) and the last NASA Discovery call ({\it Proteus} -- \citealt{Proteus}).

\section{The Main Belt Comets}\label{sec:sci}

\begin{table*}
\caption{MBC candidates}
\center
\small
\begin{tabular}{p{4cm}cccclp{5cm}}
\hline
Object & $a$ & $e$ & $i$ & $T_J$ & Nature & Comments\\
\hline
133P/Elst-Pizarro & 3.161 & 0.162 & 1.387 & 3.19 & MBC & Castalia primary target. Long lasting activity; 
Themis family member. Active during 4 consecutive perihelion passes.\\
176P/LINEAR & 3.196 & 0.193 & 0.235 & 3.17 & MBC candidate & Long lasting activity. Not active at 2nd  perihelion\\
238P/Read & 3.164 & 0.253 & 1.265 & 3.15 & MBC & Castalia secondary target. Long lasting activity. Active during 3 consecutive perihelia.\\
259P/Garradd & 2.728 & 0.341 & 15.899 & 3.22 & ? & Orbit not stable; probably recently arrived\\
288P/2006~VW139 & 3.051 & 0.199 & 3.239 & 3.20 & MBC  & Long lasting activity; Themis family member. Repeated activity*.\\
354P/LINEAR & 2.290 & 0.125 & 5.257 & 3.59 & Active asteroid & Short, impulsive activity. Probably an impact, possibly rotational disruption\\
331P/Gibbs & 3.005 & 0.041 & 9.738 & 3.27 & Active asteroid & Short, impulsive activity Impact\\
P/2012 T1 (PANSTARRS) & 3.150 & 0.237 & 11.057 & 3.18 & MBC candidate & Long lasting activity; member of Lixiaohua family\\
596 Scheila & 2.926 & 0.165 & 14.662 & 3.21 & Active asteroid & Short, impulsive activity; probably an impact\\
324P/La Sagra & 3.095 & 0.154 & 21.419 & 3.20 & MBC & Long lasting activity. Active at 2nd perihelion.\\
311P/PanSTARRS & 2.189 & 0.115 & 4.969 & 3.66 & Active asteroid & Activity over a long period, but as very short bursts: Probably rotational shedding.\\
P/2013 R3 (Catalina-PANSTARRS) & 3.036 & 0.273 & 0.899 & 3.18 & Active asteroid & Fragmentation in at least 4 pieces. Rotational disruption?\\
233P/La Sagra & 3.032 & 0.411 & 11.279 & 3.08 & ? & Poorly studied.\\
313P/Gibbs & 3.155 & 0.242 & 10.967 & 3.13 & MBC & Discovered active; Seen active at 2003 perihelion (pre-discovery), not observed 2009. Member of Lixiaohua family\\
62412 (2000~SY178) & 3.146 & 0.090 & 4.765 & 3.20 &  ? & Fast rotator, long-lived activity. Hygiea family.\\
P/2015 X6 (PANSTARRS) & 2.755 & 0.170 & 4.558 & 3.32 & MBC candidate & Long lasting activity\\
P/2016 G1 (PANSTARRS) & 2.583 & 0.210 & 10.968 & 3.37 & Active asteroid & Short disruptive event\\
P/2016 J1 (PANSTARRS)* & 3.172 & 0.228 & 14.330 & 3.11 & MBC candidate & Split, both components show long lasting activity.\\
\hline
\end{tabular}
\smallskip
\newline
Orbit parameters: $a$ = semi-major axis; $e$ = eccentricity; $i$ = inclination; $T_J$ = Tisserand parameter with respect to Jupiter.\newline
* Discovery / confirmation after M5 proposal submitted.
\label{tab:MBClist}
\end{table*}

\subsection{MBCs as a population in the planetary system}
The traditional distinction between inactive asteroids and active comets has recently been blurred by the discovery of MBCs. The first MBC discovered, 133P/Elst-Pizarro, was found in 1996. Its strange appearance, with an apparently comet-like tail but an asteroidal orbit, was a unique curiosity that attracted some debate  \citep{Boehnardt98}. It was variously described as a comet that had found some unusual route to a home in the main asteroid belt, from here on the Main Belt, or the result of an impact. MBCs were first identified as a population following the discovery of two other similar objects in the outer Main Belt   \citep{Hsieh+Jewitt06}. While a single object could have been explained as a fluke, these discoveries showed that there must be a significant population of comets in the Main Belt -- more than could have found their way there from known comet reservoirs by complex gravitational interactions. This discovery meant that a new, and entirely unexpected, population of comets had been identified, which has profound implications for the Solar System's formation, and its evolution since. The implied presence of water in the Main Belt, within the `snow line' where it was not thought possible for ice to form in the protoplanetary disc, means that either icy bodies could form closer to the Sun than was thought, and our understanding of temperatures or physics within the disc is incorrect, or that significant migration of planetesimals must have taken place early in the Solar System's history.

\begin{figure*}
\begin{center}
\includegraphics*[width=1.98\columnwidth]{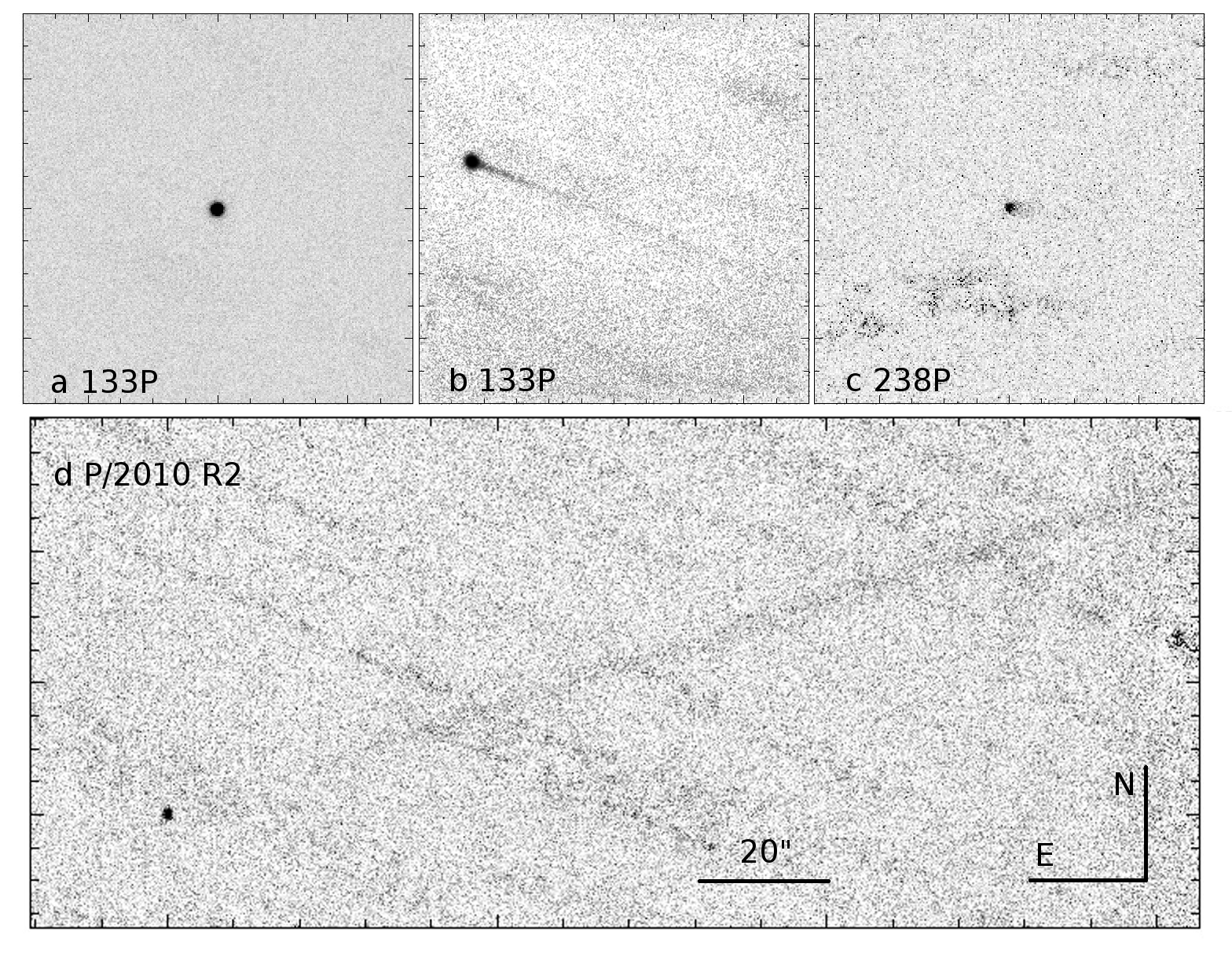}
\end{center}
\caption{MBC 133P, inactive on 2011-03-05 (a) and active on 2013-09-04 (b); 238P, showing faint activity on 2011-12-23 (c); 324P with a long, faint tail on 2011-12-23. All images were obtained through the R filter with the ESO 3.6m NTT  on La Silla. The orientation and scale are the same in all panels.}
\label{fig:133Pimage}
\end{figure*}

\begin{figure}
\begin{center}
\includegraphics*[width=\columnwidth]{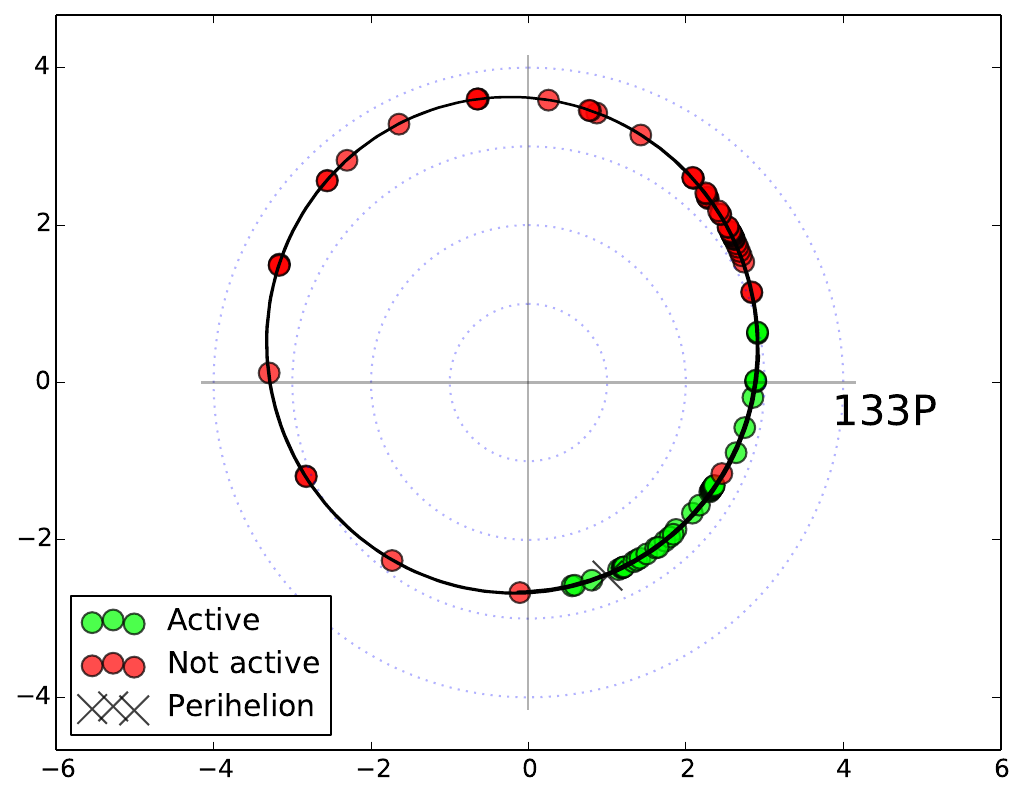}
\end{center}
\caption{This figure shows the orbit of 133P (looking down on the solar system, with x-y position in AU), indicating when it has been seen active.}
\label{fig:133Porbit}
\end{figure}

Searches for active asteroids and MBCs have accelerated since this discovery, both dedicated surveys and careful inspection of data taken from other wide field surveys. Technology improvements in asteroid searches over the past decade have contributed to many discoveries. There are currently 18 active objects known (see Table \ref{tab:MBClist}). The picture has however been complicated -- some of these objects with comet-like appearance have been shown to be short-lived debris clouds from collisions, while others seem to be genuine comets (with activity driven by the sublimation of ices). There are various methods by which an asteroid can lose material (i.e. develop a tail or coma): Sublimation of ices, impact ejection, rotational instability, electrostatic forces, thermal fracture, thermal dehydration, shock dehydration, and radiation pressure sweeping (see \cite{Jewitt12review} and \cite{Jewitt-asteroidsIV} for reviews); but the evidence that at least some of these are genuine, sublimating comets is very strong. The strongest evidence for sublimation is found for those objects (133P/Elst-Pizarro, 238P/Read, 288P/2006~VW139, 313P/Gibbs, and 324P/La Sagra) that have been observed to repeat their activity during multiple orbit passages; each time showing activity when nearest the Sun, but showing none when further away, just like `normal' comets (fig.~\ref{fig:133Porbit}). We classify objects with strong evidence of sublimation-driven activity as MBCs, to distinguish them from other active asteroids. See fig.~\ref{fig:133Pimage} for illustrations of three of the main MBCs in various states.
In establishing the MBCs as a new population, \cite{Hsieh+Jewitt06} estimated this population to number $\sim$15-150 `active' km-scale bodies at any given time, i.e. showing activity at some point in their orbit. This implies that the underlying population of `dormant' MBCs -- ice bearing bodies in the outer Main Belt that could be triggered into activity -- must be much larger. The most recent estimate by \cite{Hsieh2015b} gives a population of 140-230 active MBCs, based on Pan-STARRS survey statistics (which included 2 confirmed MBC detections).

\begin{figure}
\begin{center}
\includegraphics*[width=\columnwidth]{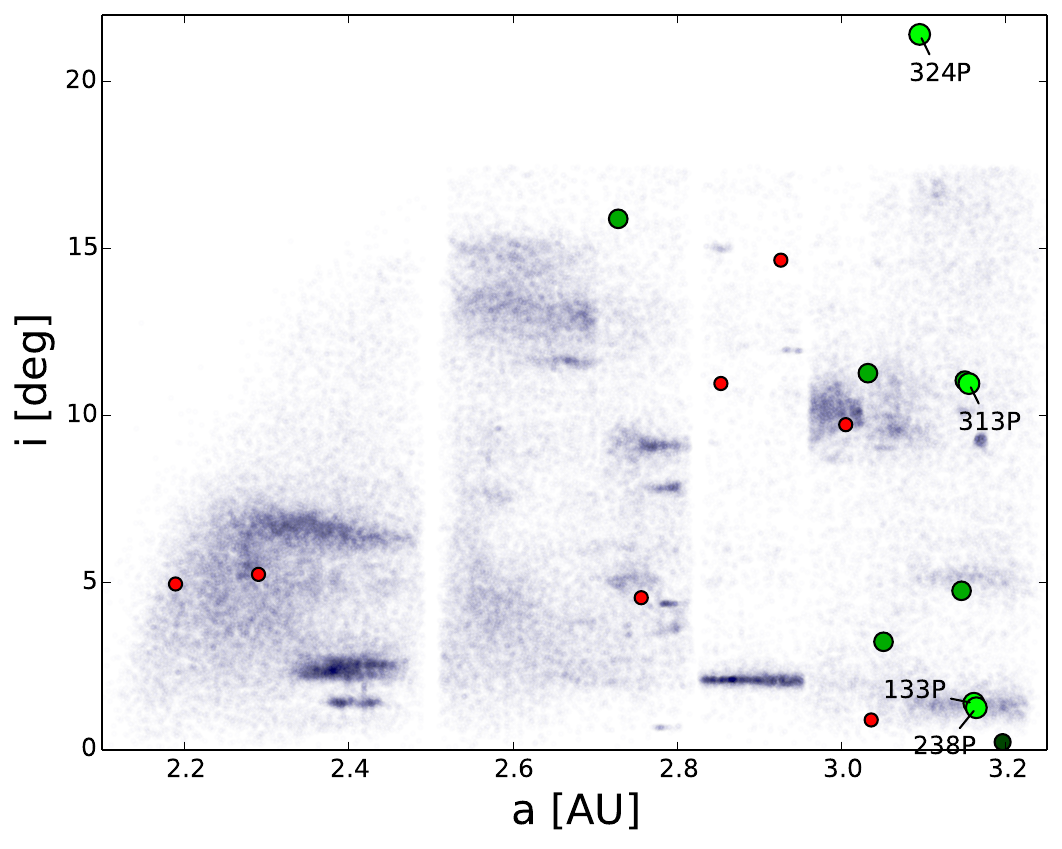}
\end{center}
\caption{Orbital elements (semi-major axis $a$ and inclination $i$) for asteroids (points), MBC candidates (green) and other active asteroids (red). The lighter green dots are the confirmed (repeating) MBCs.}
\label{fig:MBCorbits}
\end{figure}

\subsection{MBC orbit dynamics and MBC origin} 
An attempt to address the origin of MBCs has been made on dynamical grounds, but this remains an open question. Numerical simulations rule out recent capture of comets ejected from the Kuiper Belt, since the transfer and particularly the insertion from the Solar System's outer region into the Main Belt cannot be achieved efficiently for the Solar System's current configuration \citep{Levison2009}. In addition, MBCs' orbital dynamics in the Main Belt appear to be long lasting, and are stable over the age of the planetary system \citep{Haghighipour09_MBC}. \cite{Hsieh2016} demonstrated that under certain circumstances, Jupiter-family comets (JFCs) could be captured temporarily into main-belt orbits through a series of interactions with the terrestrial planets, though such captured objects would be expected to retain orbits with both high eccentricities and high inclinations.  As such, the low-inclination {\it Castalia} targets 133P and 238P are still considered likely to be native to the asteroid belt.
Hence, MBCs' origin in the Main Belt, or at least their early insertion into it, are strongly favoured scenarios. MBCs are therefore interesting bodies to test Solar System formation and evolution scenarios, such as the recent `Grand Tack' model \citep{grand-tack}, as an uniquely accessible population of icy bodies native to their current location.

\subsection{Activation scenario: impact trigger} 
For ice sublimation to be a viable activity scenario, ice must be present today. MBCs have spent between 3.9 and 4.6 Gyr in the Main Belt. The sublimation loss of exposed ice is too rapid to be sustainable over such a long-period relatively close to the Sun \citep{Sch08}. The question is whether ices can survive for billions of years if buried below the surface of these bodies, and then be a plausible driver of the observed cometary activity, as in classical comets. Thermal modelling showed that crystalline water ice may survive in MBCs, at depths from $\sim$50-150m over the entire lifetime of the Solar System at 133P's heliocentric distance \citep{PrialnikRosenberg09}, though if a MBC was produced as a fragment of a recent family-forming disruption of a larger asteroid, ice could be located much closer to the surface. These simulations also suggest that other volatile ices typically found in normal comets are very much depleted in MBCs.

To bring this buried ice close enough to the surface to drive activity, a triggering mechanism must be considered. This must have acted recently to explain the observed activity because exposed, dirty water ice located at the sub-solar point of a MBC at 2.4-2.9 AU will sublimate and recede at $\sim$1 m/yr. Given the km-scale sizes of currently known MBCs, these objects' active lifetimes once sublimation begins must be $<$1000 yr \citep{Hsieh+Jewitt06}.
The most likely triggering mechanism proposed so far is the impact of a smaller (metre-sized) object able to penetrate the insulating inactive layers and expose deeply buried ice to the Sun's heat. The exposed ice would then sustain the observed activity by insolation-driven sublimation, ejecting dust into the coma and tail as in normal comets. Recent thermal modelling suggests that ice exposure is not even strictly necessary to activate MBCs: even a very small impact could trigger the activity, bringing the Sun's heat closer to ice-rich layers still buried under a shallow mantle layer \citep{Capria12}. This hypothesis agrees with the present-day Main Belt impact frequency \citep{Hsieh09c, Capria12}.

\subsection{MBCs, water, and life on Earth} 
\subsubsection{Direct versus indirect detection of water}
In traditional comets the presence of gas, including its isotopy, is directly detected via spectroscopy of emission lines from the fluorescence of various coma gas species interacting with solar UV radiation (e.g. \citealt{Bockelee-Morvan2004}). These gases can then be used to infer the presence of various parent volatile ices in the nucleus. As MBCs are very faint comets, there is little chance of ground-based telescopes detecting any but the strongest emission lines, let alone measuring isotopic ratios. Even sensitive searches using the {\it Herschel Space Telescope} produced only upper limits \citep{ORourke2013}. 
Unfortunately, all of the available upper limits are consistent with the predicted production rates required to drive the observed dust production, and are therefore not constraining on theory -- water ice sublimation remains by far the most likely explanation for repeating MBC activity, but even for the most active MBCs is at a level below that detectable with current telescope technology \citep{Snodgrass-T1,Snodgrass-AARv}. Positive confirmation of water sublimation in MBCs should be just possible with the {\it James Webb Space Telescope} \citep{kelley16}, but certainly any detailed studies of this water (such as measuring isotope ratios), will require in situ investigation.

\subsubsection{D/H ratio and extra-terrestrial origin of Earth's water}
The prevailing hypothesis for the origin of Earth's water involves the post-accretion bombardment of the young Earth by volatile-rich planetesimals from beyond the snow line, i.e., comets from the outer Solar System or objects from the Main Belt (e.g., \citealt{Anders+Owen,OwenB95_impacts}). This is referred to as the `late veneer' model. The large ice content of comets from the outer Solar System makes them natural candidates for delivering water to the Earth. However, as we now know, asteroids from the Main Belt also contain ice and therefore could have also delivered water to the Earth, most likely at a much higher rate than comets. 

Measurement of the deuterium-to-hydrogen (D/H) isotopic ratio, hereafter D/H, provides a chemical signature of water's source region, and hence has the potential for constraining the origin of volatiles on Earth. Hydrogen and deuterium (hydrogen which also contains a neutron) were synthesised during the Big Bang \citep{Wagoner1967}, and there is no known mechanism that produces significant amounts of D in galaxies or stars \citep{Epstein1976}. The second largest Solar System deuterium reservoir after the Sun is deuterated water, HDO. D/H in water is very sensitive to the conditions in which the molecule was formed, in particular the medium's kinetic temperature. 
D/H in water is predicted to increase with heliocentric distance \citep[ for a review]{Mousis2000,Horner2007,Robert2006}. 
The D/H of the planetesimals' formation region is essentially frozen into the bodies when water ice forms. Current models indicate that D/H varies little below 5 AU, increases rapidly with heliocentric distance to 45 AU; and beyond this varies little again.

\begin{figure}
\begin{center}
\includegraphics*[width=\columnwidth,trim=400 270 400 200,clip]{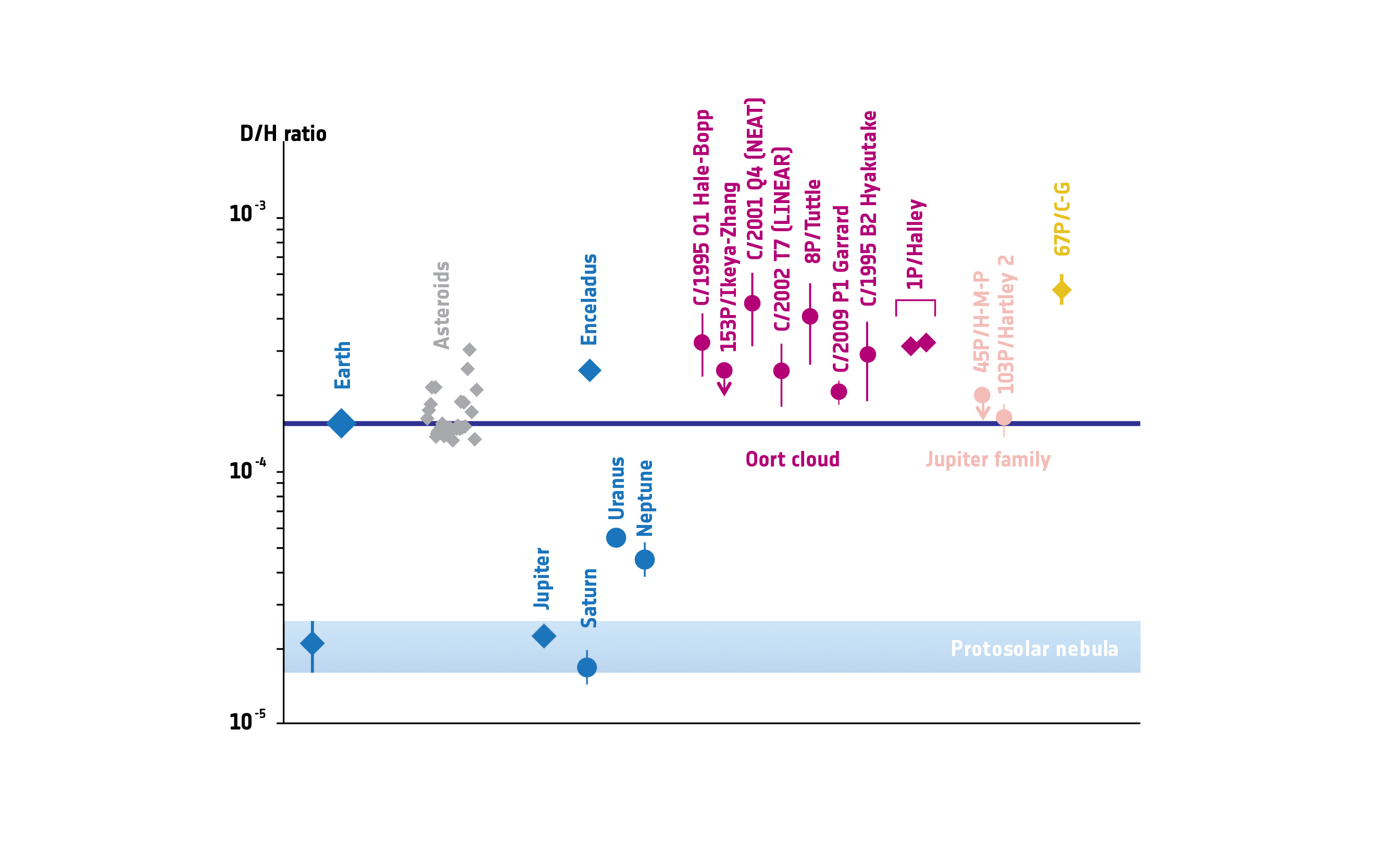}
\end{center}
\caption{D/H ratios measured in various Solar System bodies. Image \copyright ESA, based on \citet{Altwegg-DH}}
\label{fig:DH}
\end{figure}

Earth's oceans provide an interesting and paradoxical case of enhancement compared to the solar nebula value. The estimated D/H in Standard Mean Ocean Water (SMOW) is 1.6$\times10^{-4}$ \citep{Balsiger1995}, $\sim$6-7 times the primitive Sun's value, deduced from the solar wind implanted in lunar soils or from Jupiter's atmosphere \citep{Robert2000}. For comets, in situ D/H measurements were obtained at  1P/Halley by {\it Giotto} \citep{Balsiger1995,Eberhardt1995}, and from material  delivered to Earth from  81P/Wild by {\it Stardust}  \citep{McKeegan2006}. Halley's D/H is significantly higher than for terrestrial sea water. In contrast D/H for 81P/Wild was determined to cover a wide range, from terrestrial to Halley-like values. For very bright (i.e. highly active) comets, Earth-based measurements are possible: 
For Oort cloud comets, measurements all agree with an average D/H = 2.96 $\pm$ 0.25$\times10^{-4}$ (fig.~\ref{fig:DH}). Asteroidal material, measured in carbonaceous chondrite meteorites originating beyond 2.5 AU, has a more SMOW-like D/H = 1.4 $\pm$ 0.1 $\times10^{-4}$ \citep[e.g.][]{Marty2011}. This led to models in which asteroids must have been the major suppliers of volatiles on Earth, with $<$10\% of water being delivered by comets. 

However, the picture was complicated by {\it Herschel} observations of JFCs 103P/Hartley 2, with D/H = 1.61 $\pm$ 0.24 $\times10^{-4}$ \citep{Hartogh2011}, and 45P/Honda-Mrkos-Pajdu{\v s}{\'a}kov{\'a}, D/H $<$ 2.0$\times10^{-4}$ (3$\sigma$ upper limit -- \citealt{Lis2013}). These match SMOW, and are therefore incompatible with the scenario of small body formation and Nice models invoking relatively `gentle' planet migration \citep{Tsiganis2005, Gomes2005,Morbidelli2005}, which suggest that Oort cloud comets might have formed nearer to the Sun than JFCs (i.e. JFCs should have higher D/H than Oort cloud comets). 

The most recent twists in this tale, however, are observations by {\it Rosetta}'s ROSINA instrument at JFC 67P of D/H = (5.3 $\pm$ 0.7)$\times10^{-4}$ \citep{Altwegg-DH} and {\it Odin} observations by \cite{Biver2016} of D/H = (6.5 $\pm$ 1.6)$\times10^{-4}$ and (1.4 $\pm$ 0.4)$\times10^{-4}$ in the two Oort cloud comets, C/2012 F6 (Lemmon) and C/2014 Q2 (Lovejoy), respectively. This indicates a very heterogeneous variation in D/H of the objects in the outer Solar System and implies that the commonly used comet classifications reflect their dynamical history rather than their formation location. These results mean that Earth's water cannot simply come from the outer Solar System and returns the spotlight to the asteroid belt.

The Grand Tack model \citep{grand-tack}, Nice models invoking more catastrophic planet migration scenarios \citep{Tsiganis2005,Morbidelli2008,Pierens2014}, and work by \cite{DeMeo+Carry2014} suggest large-scale planetesimal mixing between the inner and outer Solar System. Therefore, a significant fraction of Main Belt asteroids may originate in the outer Solar System region, with different D/H. The measurement of D/H in MBCs, which are a key population to disentangle the different models of the Solar System dynamical evolution, would provide critical data to test these Solar System formation models, as well as providing the first measurement of the only remaining potential exogenic source of Earth's water. The same is true for other key volatiles such as the molecular oxygen, O$_2$, which is among the major species outgassing from the nucleus of 67P \citep{Bieler2015}. The presence of O$_2$, together with other species such as S$_2$ \citep{Calmonte2016}, indicate that parts of the ice in 67P never sublimated but remained frozen since formation in the ISM. This is also supported by the highly volatile N$_2$ found in 67P \citep{Rubin2015}. The abundance of these volatiles in MBCs are still unknown but obviously crucial to understand the formation and dynamics of our Solar System.
In situ measurement by a mission like {\it Castalia} is the only way to achieve isotopic measurements of key volatile species.

\section{Science Requirements for a MBC mission}\label{sec:req}
\begin{table*}
\caption{Scientific goals of the {\it Castalia} mission}
\centering
\begin{tabular}{lll}
\hline
Goal & Description & Objectives$^\dag$\\
\hline
A & Characterise MBCs, a new Solar System family, by in situ investigation & All\\
B & Understand the physics of MBC activity & 1-6, 9\\
C & Directly detect water in the asteroid belt & 2, 4-7\\
D & Test whether MBCs are a viable source for Earth's water & 4, 10\\
E & Use MBCs as tracers of planetary system formation and evolution & 2-5, 8-10\\
\hline
\end{tabular}
\\
$^\dag$ Objective numbers refer to the list in section \ref{sec:req}
\label{tab:goals}
\end{table*}

A space mission is the only adequate approach to address and solve the important planetary science questions related to the MBC population. Earth-based observations, including the use of existing and planned space observatories, are considered of complementary value, but will be insufficient to get even close to the answers provided by a dedicated mission. Thus, the scientific goals of {\it Castalia} are: To characterise a representative member of the MBCs, a newly-discovered Solar System family, by in situ investigation;  to understand the trigger and the physics of the MBC cometary activity; to directly detect water in the asteroid belt; to measure its D/H ratio to test whether MBCs are a viable source for Earth's water; and to use MBCs as tracers of our planetary system formation and evolution.
These scientific goals are projected into specific scientific objectives to be accomplished through measurements by {\it Castalia}'s instruments, as linked in Table \ref{tab:goals}. Below we list these objectives (in no particular order -- numbers are simply to allow cross referencing to science goals) and possible instrumentation options. We then present, in section \ref{sec:ins}, the selected payload complement for {\it Castalia} that best addresses these objectives whilst meeting mass and power constraints.   
\begin{enumerate}
\item Confirm the impact activation theory by identification of site: The impact region, with excavated sublimating material or surface cracks with gas release, has to be identified and its on-going activity has to be measured. Suitable instruments: visible, near-IR and thermal IR cameras, gas and dust analysers.
\item Map surface structure, geology and mineralogy including hydrated/organic minerals and search for ices: The surface structure and geology provide imprints from the MBC's formation and evolution. Surface dating is accomplished through cratering statistics. The MBC's surface structure might display signatures from asteroidal- and cometary-like evolution. The interpretation will benefit from comparing MBC findings with those of asteroids (e.g. {\it Dawn, Rosetta, Galileo, Hayabusa, NEAR-Shoemaker}), and of comet nuclei (e.g. {\it Rosetta, Deep Impact, Stardust}). Surface mineralogy gives complementary information in terms of the MBC's formation and evolution environment. The detection of hydrated minerals and organics compounds, and of the presence of surface ices, for instance in the MBC's active region(s), would be of particular value. Suitable instruments: visible and near-IR imagers and spectrographs, thermal-IR imagers and spectrographs, radar.   
\item Determine the elemental and molecular composition, structure and size distribution of the dust and compare it to results for comets and asteroids. Dust is considered representative of the surface material's non-volatile component and provides links to dust compounds measured in asteroids and comets, including in the samples that will have been returned to Earth from asteroids by the {\it Hayabusa 2} and {\it OSIRIS-REx} missions by the time {\it Castalia} flies.  Suitable instruments: dust counting and composition analysers complemented by visible and thermal imagers and IR spectrometers.  
\item Determine elemental and molecular abundances of volatile species (or constraining upper limits): The measurements shall identify and quantify the gaseous compounds produced by the MBC activity. Water is the expected key species, but others typically seen in comets (e.g. CO, CO$_2$), though low in amount in MBCs, are of high interest to constrain MBC evolution models. Suitable instruments: primarily mass spectrometer and gas chromatographs, then microwave and visible/near-IR spectrometers. To a lesser extent, filter imagers in reflected and thermal light.
\item Characterise or constrain the subsurface and internal structure of an MBC:  Shallow subsurface sounding provides the stratigraphic relation between surface geological units, allowing the modelling of geological structures' formation and evolution (ice layers, terraces, craters, boulders, redeposit and regolith?). It is a key point to identify and model the activation mechanism and jet formation (endogenous or exogenous origin, relation with ice layers). The deep interior provides information on the dust/ice bulk ratio and on the geological structures at global scale. It aims to better model accretion mechanism and the body's evolution. Suitable instruments are high frequency radar for high-resolution sounding of the shallow subsurface (first tens of metres) and low frequency radar for deep interior sounding, and radio science (spacecraft tracking).
\item Characterise the diurnal and orbital activity cycles of an MBC: In addition to the basic geometric and rotational parameters of the body, these cycles are controlled by the thermal environment, and the thermal characteristics of the surface and subsurface.  These, and the activity itself, are best addressed by imaging and/or spectroscopic instruments in reflected and thermal light, by gas mass spectrometers and dust analysers. 
\item Characterise the plasma environment of a weakly outgassing object and its interaction with the solar wind: This is widely unexplored; even in its earliest `low-activity' phase {\it Rosetta} measured a stronger plasma environment (by at least one order of magnitude) than expected at an MBC. Theoretical modelling predicts a plasma environment and geometry that differs considerably from that encountered by space probes so far. Suitable instruments are a plasma analyser and a magnetometer.   
\item Search for a primordial inherent magnetic field in a small Solar System body: The role of magnetic forces in the formation of planetesimals is unclear. If relevant, remnant magnetic fields on local scales may exist in small bodies and could be assessed by near-surface magnetic field measurements. The suitable instrument is a sensitive magnetometer.
\item Determine global physical properties of MBCs: mass, volume, gravity field, thermal properties: these are needed to estimate basic body parameters which are also indicative of the MBC's interior, i.e. mean density, interior mass distribution, and thermal properties. Suitable instruments: Visible/near-IR and thermal imagers as well as radar and radio science provide the required measurements and information.
\item Determine the MBC's D/H ratio and isotopic composition: Isotopic measurements will constrain the body's formation region in the Solar System. Moreover, D/H and oxygen isotope ratios of water in the MBC coma can indicate whether MBCs are a potential source for terrestrial ocean water. Suitable instrument: Neutral and ion mass spectrometer for in situ measurements of the MBC's gas environment.
\end{enumerate}

These objectives address very clearly two of the themes of the ESA Cosmic Vision 2015-2025 programme, namely Theme 1: `What are the conditions for planet formation and the emergence of life?', and Theme 2: `How does the Solar System work?', in particular sub-theme 2.3: Asteroids and Small Bodies. In situ sampling with proven instrumentation of gas and dust from a volatile-rich small body provides a cost-effective means of determining the composition of an important new type of object.
The investigation of the solar wind's interaction with a low activity body also addresses sub-theme 2.1: From the Sun to the edge of the Solar System.

On the basis of {\it Castalia}'s scientific goals, several mission types were evaluated for the chances of successful achievement of the scientific objectives. The `Orbiter' mission at a single target appears to be the simplest mission type for achieving all the science goals, being also the least hazardous and the most economical. The other mission types that were considered -- multiple target orbiter, sample return, in situ surface mission, orbiter with surface penetrator - can contribute additional science which is considered appealing, but are not expected to open new scientific pathways for a more comprehensive understanding of MBCs and their role in the planetary system. Moreover, these mission types are more complex, demanding and costly. Nevertheless, adding a simple surface science unit is seen as an attractive optional element for the mission, capable of providing complementary scientific results to a single target `Orbiter' mission. A fly-by mission with a single or multiple targets is considered challenging for achieving the science goals and objectives, particularly for the important isotopic measurements, as the total gas collection time would be short.

The following science-driven requirements are derived for the mission:
\begin{itemize}
\item To perform an orbiting mission at a MBC, allowing remote sensing and in situ measurements by scientific instruments,
\item To explore the MBC when it is active and inactive
\item To perform a low-altitude phase for the exploration of a surface active region
\end{itemize}
The mission design proposed to achieve this is described in section~\ref{sec:mission} below. The best potential target to meet {\it Castalia}'s science goals appears to be 133P/Elst-Pizarro: It is the best studied MBC and it has been reliably and predictably active around perihelion during four consecutive revolutions. Comet 238P/Read is another suitable candidate, with activity observed during three consecutive revolutions, and is a good backup option.

\section{The target: Properties of 133P/Elst-Pizarro}\label{sec:target}
As the first discovered, and therefore best studied, 133P has the most complete characterisation. Although MBCs are very faint comets, their dust tails can be easily imaged with medium sized telescopes, and 133P's activity cycle has therefore been well mapped and found to be predictably repetitive \citep{Hsieh10,Hsieh2013CBET}. It shows activity (detectable tail, see  Figure \ref{fig:133Pimage}) for $\sim$1 year out of its $\sim$6 yr orbital period, starting around one month before perihelion. When inactive, the nucleus is faint, yet bright enough to be studied using large telescopes. Its effective radius is $\sim$1.9 km \citep{Hsieh09b}, and has a short spin period for a comet of 3.471 hr \citep{Hsieh04}. This period reveals more about the nucleus: like all $\sim$km scale bodies, 133P is expected to be a loosely bound `rubble pile', rather than a solid lump of rock or ice. Hence only gravity binds it, which can be balanced against the spin rate to limit its bulk density. For typical asteroids, there is a clear rotation period barrier of 2.2 hr that corresponds to a density $\sim$3 g cm$^{-3}$ \citep{Pravec02}, while for Jupiter family comets a lower density of $\sim$0.6 g cm$^{-3}$ is implied \citep{Snodgrass08}. 133P rotates faster than other comets, which requires a higher density, i.e. a higher rock:ice ratio, and/or lower porosity. The 3.47 hr period however implies a low minimum density, lower than most asteroids, allowing in turn a value between comets and asteroids for this cross-over population. 133P's average surface reflectance properties are also well studied, with visible and near-IR spectroscopy performed to search for mineral signatures. 133P's spectrum is featureless, like other comets, but with a different slope -- it is a relatively neutral/blue object, while comets tend to be quite red \citep{Licandro11b}. However, the nucleus's 5\% albedo in the $R$-band albedo is similar to other comets \citep{Hsieh09b}. Other MBC candidates are less well observed, but similar to 133P where data exist, e.g. 176P/LINEAR has a similar albedo and spectrum \citep{Hsieh09b,Licandro11b}. 

\section{Castalia mission}\label{sec:mission}

\begin{figure}
\begin{center}
\includegraphics*[width=\columnwidth]{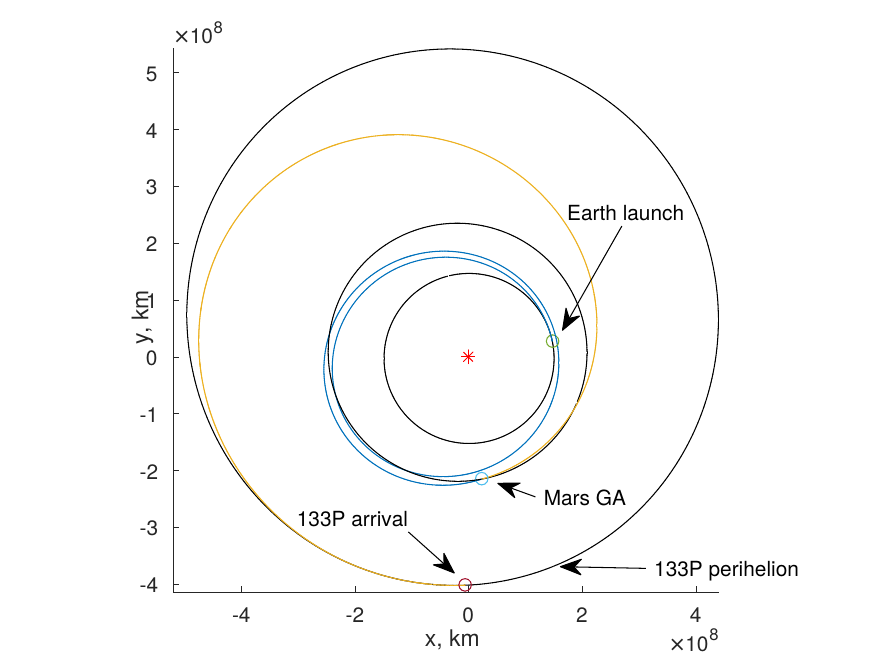}
\end{center}
\caption{Baseline EP trajectory to reach 133P, departing 2028 and arriving 2035 (see table \ref{tab:transfer}).}
\label{fig:transfer}
\end{figure}

\begin{figure*}
\begin{center}
\includegraphics*[width=\textwidth]{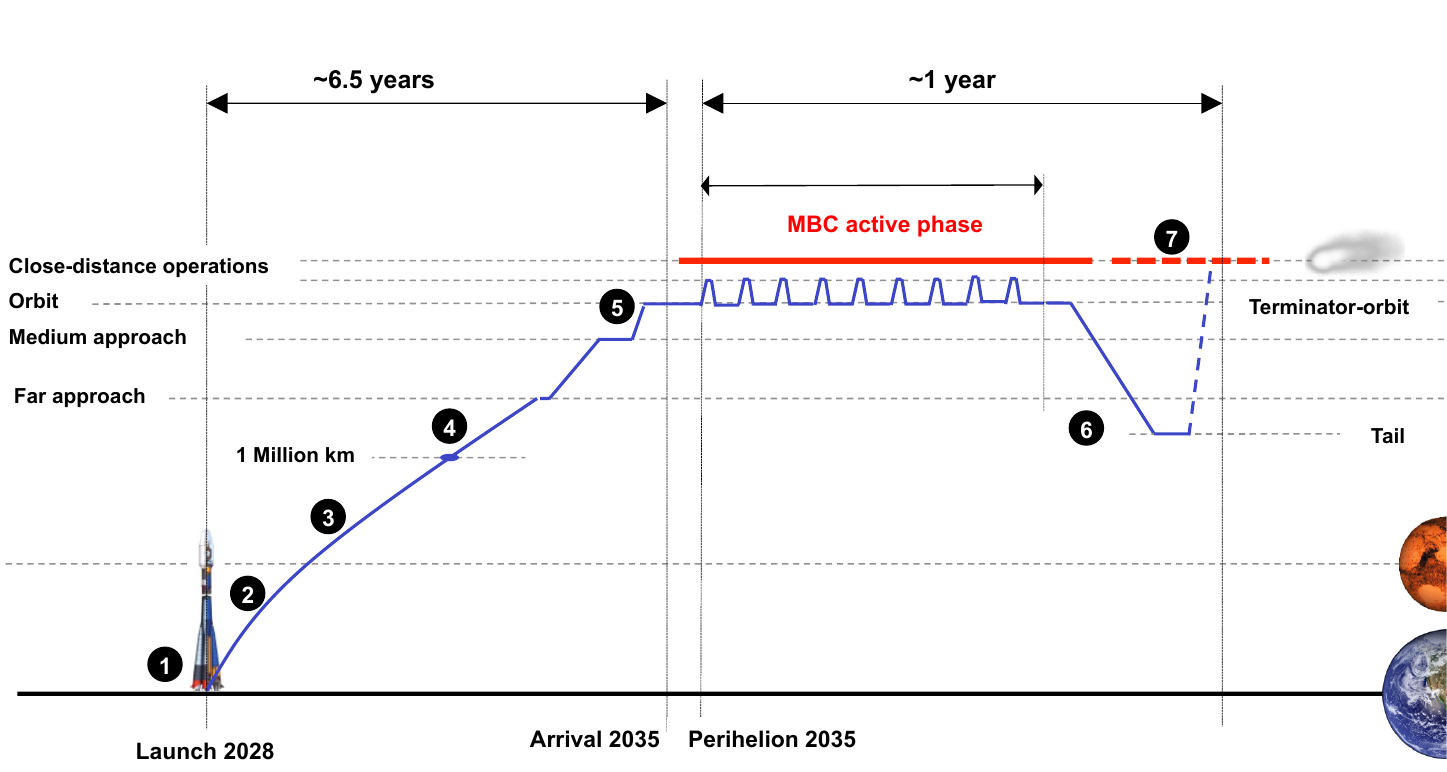}
\end{center}
\caption{Outline of mission architecture: 1. Launch; 2. Mars GA; 3. Optional cruise flybys; 4. Target acquisition; 5. MBC phase; 6. Tail excursion; 7. Optional landing (EOM).}
\label{fig:architecture}
\end{figure*}

The launcher is assumed to be the Ariane 6.2 from Kourou, with the performance values of the Soyuz Fregat 2.1B used for calculations as the Ariane 6.2 parameters are not yet available. This represents the assumed `worst-case' performance for a European medium class launch vehicle in the late 2020s, and can provide a characteristic energy ($C_3$) value of 11.5 km$^2$ s$^{-2}$ for a spacecraft wet mass of 1611 kg. As part of the iterative mission and system design, transfers have been computed considering electric propulsion (EP) and chemical propulsion (CP), with and without gravity assist(s) (GA). A thruster performance similar to the available RIT-22 or T-6 engines was assumed ($I_{sp} = 4200$ s, $T_{max} = 160$ mN) for EP. CP options were computed with a multi-gravity assist trajectory model using patched conic approximation and one deep space manoeuvre in each interplanetary leg \citep{Ceriotti2010}. The following gravity assist sequences were considered: EEMA\footnote{Earth(E), Mars(M), Asteroid/MBC(A), Venus(V)}, EEMMA, EMMA, EVEA and EVMA. The problem was optimised for the total $\Delta v$ (weighted sum of the of launch, manoeuvres, and arrival). A genetic algorithm \citep{Goldberg89} was run one hundred times for each case, and feasible solutions were further refined using sequential quadratic programming \citep{Byrd00}. For EP options, due to constraints on the total flight time, only the following GA sequences were considered: EA (direct) and EMA. EP trajectories were first  modelled using a shape-based approach, using a linear-trigonomeric shape \citep{DePascale06}. Again a genetic algorithm was used to minimise the propellant mass, while satisfying the temporal and spacial fitting of the trajectory. Feasible solutions were then used as first guess to run an optimal control solver using collocation on adaptive Legendre-Gauss-Radau (LGR) nodes (GPOPS-II, \citealt{Patterson14}). The subsequent nonlinear programming problem was solved with IPOPT \citep{Wachter06}. EP direct transfers and with one GA of Mars resulted in viable options, while CP options were too expensive in terms of propellant mass.

A rendezvous with 133P before its perihelion passage is desirable as the expected active period begins around one month pre-perihelion. It was found that the recommended launch window for ESA Cosmic Vision M5  missions, between 2029/01/01 and 2030/12/31, is particularly unfavourable, as it is not possible to reach the MBC for its 2035 perihelion passage, requiring a long transfer ($\sim$10 years) for the 2040 perihelion. For this reason, earlier launch opportunities were investigated. A quick and efficient transfer ($\sim$6.5 years, via Mars GA) launching in late 2028 was found for the 2035 perihelion, and selected as the baseline (Figure~\ref{fig:transfer} and Table~\ref{tab:transfer}). In this specific solution the arrival is 3 months before perihelion, but the transfer allows the arrival date to be moved by a few months without additional cost, due to the flexibility of EP trajectories. Table~\ref{tab:transfer} also shows the details of a backup option that launches within ESA's recommended M5 boundary conditions, for the 2040 perihelion.

\begin{table}
	\label{tab:transfer}
	\caption{Transfer options to 133P.}
	\centering
	\begin{tabular}{p{4cm}ll}
	\hline
		                      & Baseline   & Backup \\
	\hline
		Sequence              & EMA        & EMA \\
		Launch                & 2028/10/03 & 2030/09/29 \\
		Mars GA               & 2031/08/09 & 2033/06/08 \\
		Arrival               & 2035/05/01 & 2040/11/17 \\
        Duration, yrs         &  6.57      & 10.1 \\
		Max thrust, mN        &   160      & 160 \\
        Propellant mass, kg   & 223.5      & 227.9\\
        Possible dry mass, kg &  1387      & 1383 \\
        Time to perihelion at arrival, d & 93 & 120 \\
	\hline
	\end{tabular}
\end{table}

The transfer phase may include possible target of opportunity fly-by encounters with minor bodies, which would be valuable for both scientific and instrument calibration reasons.

During 2017 the M5 evaluation process has already been delayed, so possible trajectories with even later launch dates (within the year 2031) were investigated. The 2035 133P perihelion becomes impossible to reach; for 133P the best 2031 launch (January) gives a 9.9 year cruise with a Mars GA. Although this is a long transfer, it is relatively cheap in terms of fuel mass (235 kg), and spends a long time in the asteroid belt during the cruise, raising the probability that interesting fly-by targets can be found en-route. Shorter transfers to 238P (6.8 years), 313P (4.8 years) and 324P (5.8 years) are possible for a 2031 launch, all arriving a few months before their respective perihelion dates, although with significantly higher fuel mass required for the latter two ($\sim$ 360 kg), which have high inclinations. Further refinement of trajectory options will be possible once final launch dates are known -- in all cases the advantage of EP being relatively flexible gives a range of possible arrival dates.

At a distance of 1,000,000 km from the MBC, the Near Target Phase starts. About 1 month will be used for optical navigation with {\it Castalia}'s science and/or navigation cameras, to acquire the target and prepare for matching orbits. The following Far Approach Phase ends at a distance of about 1000 km from target. It includes the insertion into the heliocentric orbit of the target. There, intermediate science activities will be performed. The Medium Approach Phase reduces the distance between spacecraft and target to about 100 km, where global characterisation activities will be performed. A pyramidal orbit approach, as used by {\it Rosetta}, can be used to improve the global characterisation. The spacecraft stays outside the sphere of influence of the MBC during the entire Medium Approach Phase. A period of 40 days plus transition time but including the transmission time is sufficient to acquire all necessary scientific data.

In the next major phase, the MBC phase, the spacecraft approaches the object and is inserted into an orbit around the target (Close Approach Phase). In this orbit, which has about 20 km distance to the object, the target will be studied until the onset of the activity. We note that the MBCs are of a comparable size and mass to 67P, but with orders of magnitude lower gas production even in their most active phase, making {\it Castalia} a much simpler task than {\it Rosetta} for trajectory planners. 

After onset of activity the spacecraft will approach the surface several times, reaching an altitude of $\sim$5 km, `hover' there for some time, retreat and transmit the gathered data to ground. The so-called `hover' mode is a period under automated spacecraft control (see section \ref{sec:spacecraft}), where periods of quiet spacecraft data collection alternate with thruster firings to maintain altitude within a set altitude range.
While hovering, the active area of the MBC will transit below the spacecraft and in situ measurements of the gas and dust can be performed.
Although the measurement time for one transit is short (depending on rotation period, distance to surface, and size of the active region) the desired 100h of sampling time can be achieved by hovering continuously throughout various transits and performing several hovering cycles. While clearly presenting a technical challenge, hovering was considered to be the most suitable method to achieve the necessary sampling of the dust and gas close to the MBC surface.

The download of all gathered data takes place between the hovering cycles in the $\sim$20km orbit within a recurring dedicated communication mode. This simplifies the challenging hovering phase by reducing the parallel activities on-board and eliminates the need for a steerable antenna. Furthermore, this approach ensures a timely science data return and minimum risk of data loss due to any system error. 

The operational phases in the MBC's vicinity and the related spacecraft trajectory are also visualised in Figure \ref{fig:architecture}. 
As add-on to the mission a Tail Phase (10,000 km distance) can be implemented to enhance the scientific return of the mission. The data to be gathered in this phase has been judged as add-on and is not mandatory. 

At the end of {\it Castalia}'s nominal mission, with all science goals achieved, as the spacecraft poses no risks in a stable heliocentric orbit far from Earth and other planets, and with no planetary protection issues, its operation can simply be ended by ceasing communications. 
We do note, however, the possibility of attempting a landing on the MBC at the end of mission, as achieved by {\it NEAR/Shoe-maker} at asteroid Eros and {\it Rosetta} at 67P.  A targeted landing near the active region could have a particularly high science return. 

\section{Castalia payload}\label{sec:ins}

\begin{table*}
\caption{Instrument summary}
\centering
\begin{tabular}{p{5cm}lllp{2cm}p{5cm}}
\hline
Instrument  & Mass  & Power & TRL & Pointing & Heritage\\
(Name, Purpose) & (kg) & (W) \\
\hline
\multicolumn{6}{c}{Surface reconnaissance (remote sensing) package} \\
MBCCAM, vis./near-IR imager & 20 & 17-30 & $>$5 & Nadir, limb, dust & DAWN FC, Rosetta OSIRIS WAC\\
  TMC, thermal IR imager & 4 & 10-12 & 8 & Nadir, limb & UKTechDemoSat,\\
\hline
\multicolumn{6}{c}{Body interior package} \\ 
SOURCE, deep radar & 20.2 & 34.7 & $\ge$6 & Nadir & MRO Sharad, MEx MARSIS, Rosetta CONSERT\\
  SSR, shallow radar & 2.4 & 90-137 & $>$5.6 & Nadir & ExoMars WISDOM,  CONSERT \\
 Radio Science & n/a & n/a & 9 & n/a  & Rosetta RSI\\
\hline
\multicolumn{6}{c}{Material and composition (in situ) package}
 \\ 
 CAMS, mass spectrometer & 6.44 & 14.03 & 7 - 9 & Nadir & Rosetta ROSINA\\
 -- inc. COUCH,  gas concentrator & 1.5 & 20 & 6 - 9 & Nadir & Philae Ptolemy\\
  DIDIMA, combined dust detector \& composition & 23 & 27 & 5 - 6 & Nadir & Rosetta GIADA and COSIMA\\
\hline
\multicolumn{6}{c}{Plasma environment package} \\ 
MAG, magnetometer & 4.5 & 5 & 5 - 9  & None & CINEMA\\
  ChAPS, plasma package & 0.65 & 1 & 5 - 6 & Various & TechDemoSat, Solar Orbiter\\
\hline
\end{tabular}
\label{tab:ins}
\end{table*}

For mass and power reasons, not all suitable experiments can be included in {\it Castalia}'s scientific payload. To attain the mission's scientific goals and objectives, the scientific instruments presented in Table \ref{tab:ins} are proposed. These are grouped into four instrument packages, each containing separately contributed instrument sub-systems based on previous instruments (high Technology Readiness Level [TRL]). These are:
\begin{itemize}
\item MBC surface reconnaissance package: visible, near-IR and thermal-IR cameras
\item MBC body interior package: radars and radio science
\item MBC material and composition package: mass spectrometers for gas and dust, dust counter
\item MBC plasma environment package: plasma instruments \& magnetometer
\end{itemize}
The grouping is based on complementary science and operation needs: The surface reconnaissance (remote sensing) package will largely drive operations when in the higher bound orbits, with off pointing possible; the composition (in situ) package will drive the requirements for the low altitude (hovering) operations; the interior package will operate in both modes, but primarily will use nadir pointing; the plasma package will only rarely have specific orbit or pointing requirements. The interface with ESA is therefore simplified, while further mass and power savings will also be possible, e.g. by sharing data processing units. The baseline dust instrument is a combined detector and analyser that merges elements of {\it Rosetta}'s GIADA and COSIMA instruments. The cameras, radars, mass spectrometer and dust composition analyser are identified as key instruments for the main mission goals and objectives. In the following sub-sections we describe each proposed instrument in detail. 

\subsection{MBC Surface reconnaissance package}
The surface reconnaissance package comprises remote sensing instruments to map the nucleus surface from large and intermediate distance (approach and orbiting phases), and obtain high resolution imaging of the area under the spacecraft in close hovering mode. It is based on multi-filter imaging systems with different detectors and optics appropriate for different wavelength ranges.

\subsubsection{Visible/Near-IR Imager: MBCCAM} 

The science objectives for {\it Castalia}'s camera package MBCCAM are:
\begin{itemize}
\item Determination of global physical parameters of the MBC body
\item Topography characterisation for geological context analysis including surface terrain models
\item Characterisation of surface morphology and body geology including age estimations and surface layering 
\item Surface mineralogy assessment, including presence of ices, organics and stony materials
\item Characterisation of light scattering properties of the surface and its granularity
\item Identification of surface activity of dust and gas emission
\item Assessment of MBC activity trigger scenario through the identification of the impact structure and possible surface ices 
\item Characterisation of the gas and dust environment in the MBC coma and tail
\item Search for heavier dust grains, boulders and satellites in the MBC's neighbourhood
\end{itemize}

To perform high-resolution surface studies and to enable detection of the potentially small ($\sim$10 m wide) ice spot for MBC activity, a spatial resolution of $\sim$5 m from 20 km distance is considered adequate. Global surface coverage (whole body per exposure or small exposure set) is required for activity studies. The detection of surface ices, organics and minerals as well as coma gas and dust needs wavelength coverage from the short-wave visible to the near-IR (minimum 2.5 $\mu$m, aim is 3.5-4.0 $\mu$m). Filter imaging of specific wavelength regions for compound detection as well as for continuum characterisation are required. The camera's photometric accuracy should aim for at least 2\%; astrometric corrections should reduce distortions to a negligible level. The scientific camera package for {\it Castalia} contains two cameras, one for the high resolution imaging in the visible (MBCCAM-v) and one for wide-angle imaging in the near-IR wavelength range (MBCCAM-i).

\begin{table}
\caption{MBCCAM parameters}
\begin{tabular}{p{33mm}ll}
\hline
 & MBCCAM-v & MBCCAM-i\\
\hline
 Focal length (mm) & 150 & 135\\
F number & 7.5 & 5.6\\
Field of view (deg) & 5.5 x 5.5 & 11.4 x 12.1\\
Spectral range (nm) & 380-1100 & 900-2500/3500\\
Spectral transmission camera & $>$75\% & $>$75\%\\
Filter positions & 8 & 8\\
Detector pixels & 1024x1024x2 & 1024 x 1024\\
Pixel size ($\mu$m) & $\sim$15 & $\sim$30\\
Surface coverage of pixel at 20 km (m) & 1.9 & 4\\
AD (bit) & 16 & 16\\
estimated Power (W) & 17 & 30\\
Mass (kg) & 6 & 14\\
TRL & $>$5 & $>$5\\
In-flight heritage & Dawn FC & OSIRIS WAC\\
\hline
\end{tabular}
\label{tab:mbccam}
\end{table}

The camera concept for MBCCAM-v is based on the {\it Dawn} mission framing camera \citep{DawnFC}, while MBCCAM-i is based on the wide-angle camera  of the {\it Rosetta} OSIRIS instrument \citep{Keller-OSIRIS}. The MBCCAM-v is  a shutter-less, fixed-focus dioptric design for the visible range of 400--1100 nm. It uses a frame transfer CCD of 14 $\mu$m pixel size that supports a 1024$\times$2014 pixel field of view. The optics are optimised with $>80$\% encircled energy per pixel over the full wavelength range. The instrument consists of a stack of two major components, the camera head with the opto-mechanics, the filters, the detector and its front-end-electronics. The MBCCAM-i is based on a fixed-focus reflective design and covers a near-IR wavelength range of at least 900-2500 nm (aim: 3500 nm). It contains a mechanical shutter and a filter wheel. The $\sim$12$^\circ$ field of view is covered by a 1024$\times$1024 pixel detector. Similar to MBCCAM-v, the MBCCAM-i units are contained in a stack of two major components, the camera head and the electronics box (which can be combined with the MBCCAM-v E-box). 

The MBCCAM cameras are mounted on a common optical bench to the spacecraft structure. Camera operations uses mostly nadir pointing, but the science objectives (such as identifying active areas using limb pointed views) and the monitoring of the camera calibration (through standard star fields) also require occasional off-nadir pointing. The MBCCAM science objectives are associated with all phases of the {\it Castalia} mission at the MBC. 

\subsubsection{The Thermal Mapper for Castalia} 

\begin{figure}
\begin{center}
\includegraphics*[width=\columnwidth]{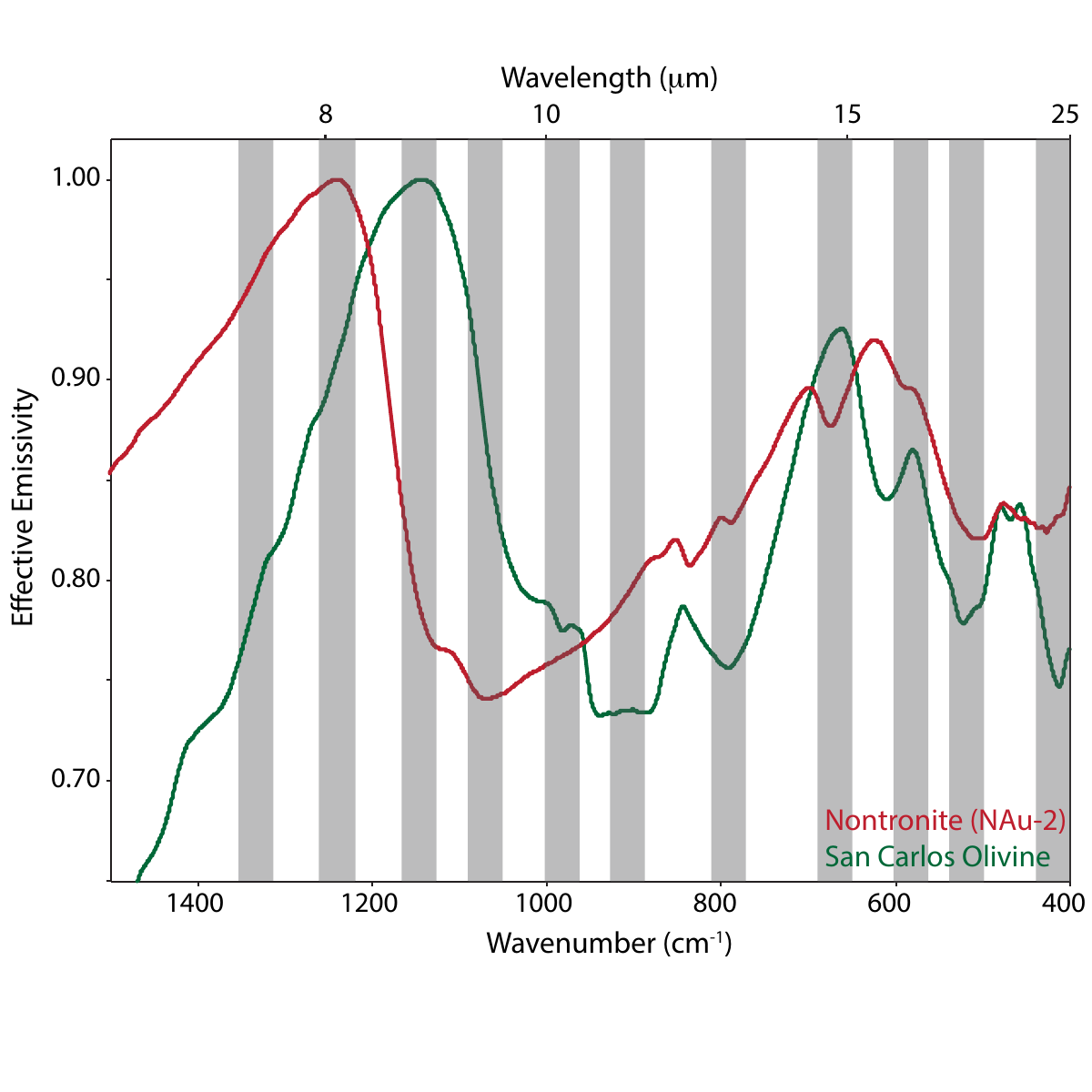}
\end{center}
\caption{Two example thermal emission spectra of olivine (anhydrous silicate mineral) and nontronite (aqueously altered  silicate mineral).  Shaded regions indicate example spectral bands for TMC that provide compositional discrimination in addition to temperature mapping.}
\label{fig:TMCspec}
\end{figure}

By mapping the MBC's diurnal temperature response, a thermal (e.g. 5 -€"- 20 $\mu$m) multi-spectral imaging instrument will provide key information on the thermal physical nature of the surface and near sub-surface (the thermal inertia) and the surface composition. 
Spatially resolved measurements of thermal inertia and composition are also vital for planning targeted observations of active regions and placing those observations into their wider geological context.  
There is significant heritage in the use of thermal infrared mapping to determine an object's surface properties, including the rich data sets produced by instruments such as the Diviner Lunar Radiometer Experiment (`Diviner', \citealt{paige2010}) on board NASA's {\it Lunar Reconnaissance Orbiter} and the Thermal Emission Imaging System (THEMIS, \citealt{Christensen2004}) on board NASA's {\it Mars Odyssey}. 
By selecting a number of narrower (e.g. $\pm$0.1 -- 0.2 $\mu$m) channels, the thermal mapping instrument is also sensitive to variations in surface composition (Figure \ref{fig:TMCspec}).

By combining near infrared remote sensing measurements (MBCCAM-i) with narrow-band thermal infrared measurements during the survey and mapping phases of the mission, the Thermal Mapper for {\it Castalia} (TMC) will help to constrain our understanding of the surface of the MBC.  This technique of combining near infrared  and thermal infrared multi-filter data has a proven heritage in providing new constraints on the composition of the lunar surface \citep{dh2014,arnold2016} and will be applied to the {\it OSIRIS-REx} \citep[e.g.][]{lauretta2015} observations of target asteroid Bennu \citep{dh2017}.

\begin{figure}
\begin{center}
\includegraphics*[width=\columnwidth]{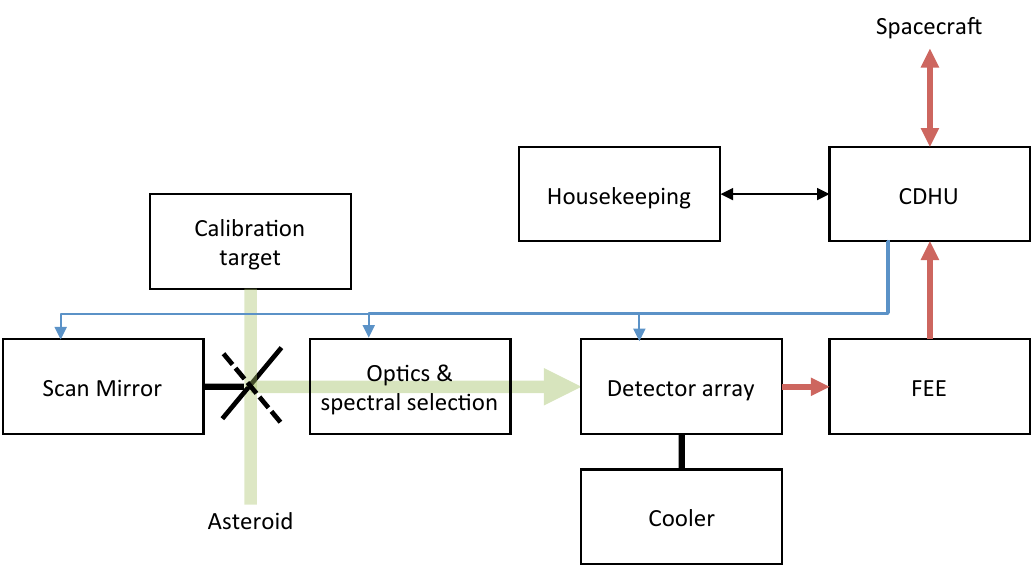}
\end{center}
\caption{TMC block diagram.}
\label{fig:TMCblock}
\end{figure}

\begin{figure}
\begin{center}
\includegraphics*[width=\columnwidth]{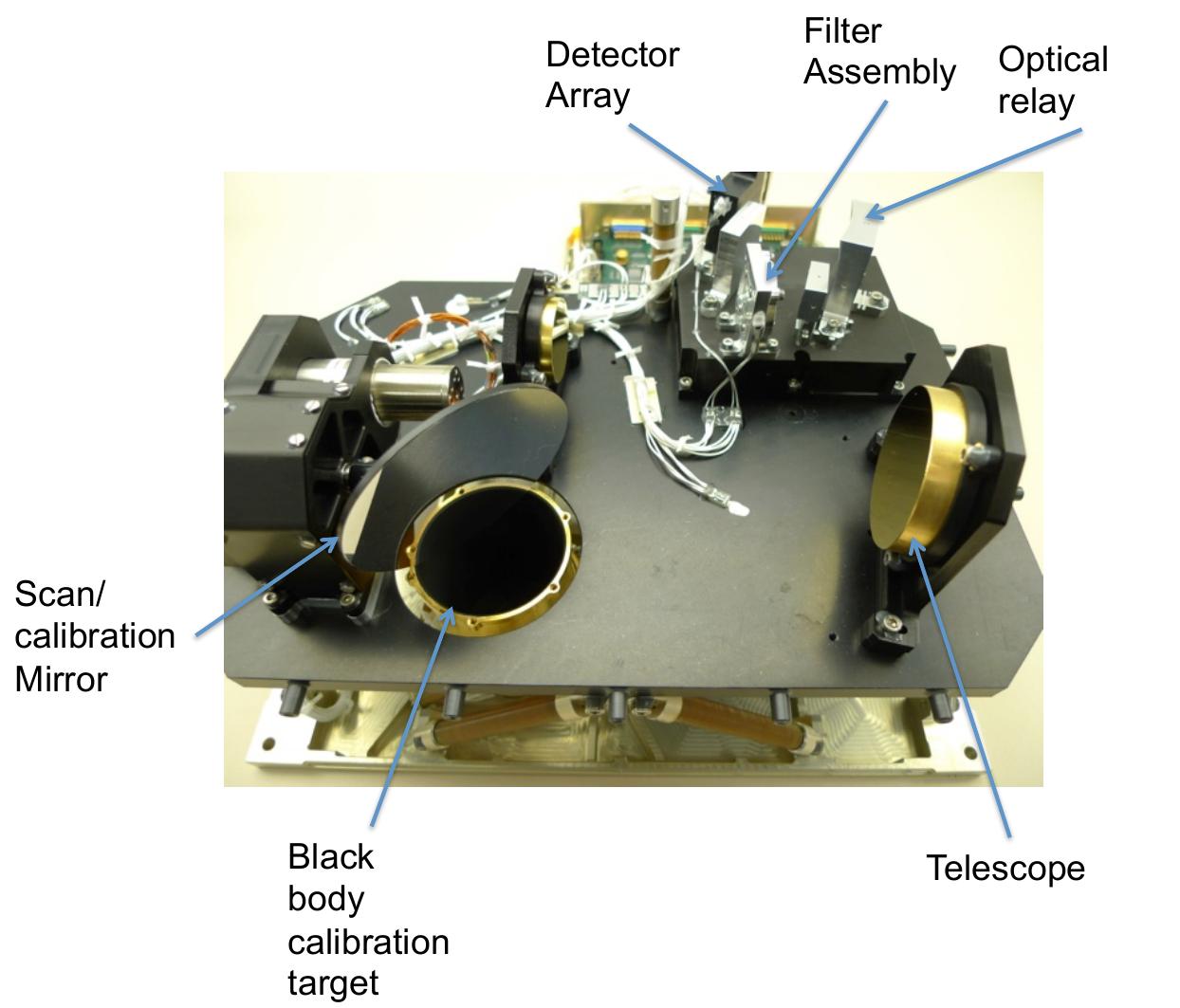}\\
\includegraphics*[width=\columnwidth]{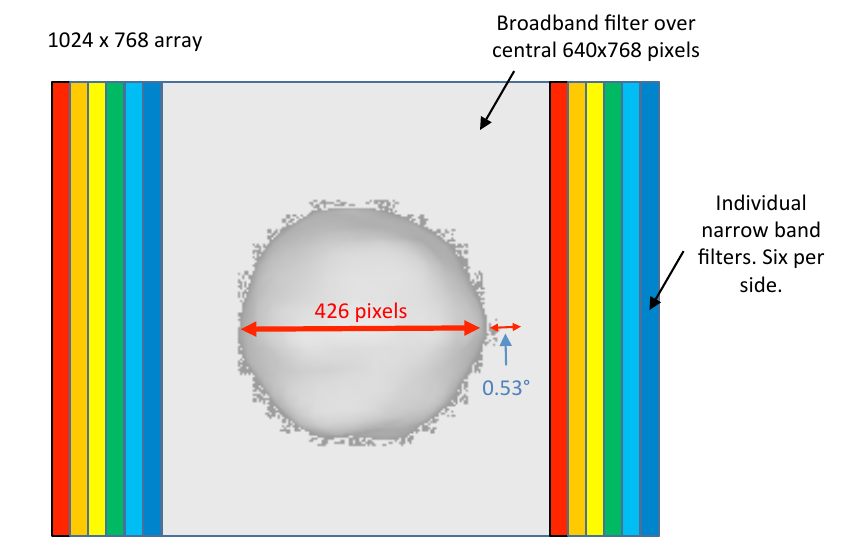}
\end{center}
\caption{(Top) The TMC heritage instrument, the Compact Modular Sounder (CMS), flight unit minus its outer cover.  CMS has approximate dimensions of 380 $\times$ 335 $\times$ 185 mm.  
(Bottom) Proposed filter and imaging detector layout, example shown during approach phase to the MBC.  The TMC will use an optimised layout, reducing the instrument size to 180 $\times$ 150 $\times$ 120 mm. Mass is 4 kg based on the CMS design.}
\label{fig:TMC}
\end{figure}

The TMC is a compact multichannel radiometer and thermal imager (Figure \ref{fig:TMCblock}) based on the Compact Modular Sounder  instrument currently flying on the UK's {\it TechDemoSat-1} spacecraft in low Earth orbit.  The TMC instrument uses a two-dimensional uncooled microbolometer detector array to provide thermal imaging of the comet (Figure \ref{fig:TMC}, bottom).    The eleven narrow-band infrared filters are mounted at an intermediate focus to improve spectral performance.  Calibration is maintained by an internal blackbody target and access to a space view and using a single scan/calibration mechanism.  Multi-spectral images are generated by push broom scanning of the target.

\begin{table}
\caption{TMC parameters}
\begin{tabular}{p{3.6cm}p{4.4cm}}
\hline
{Parameter}  & {Value/Description} \\
\hline
Spectral Range	& 	 5 -- 20 $\mu$m\\
Spectral resolution	 &	0.2 $\mu$m\\
Number of Channels &	 11\\
Temperature accuracy	 &	$\pm$5 ($\pm$1 goal) K\\
Temperature range		& 100 --€" 200 K\\
Emissivity accuracy	&	1 \%\\
Type of optics	 &	Aspheric Al Mirrors, multilayer interference filters\\
Focal number	 	& \#1.7\\
Aperture diameter  &	50 mm\\
Type of detectors		& Uncooled 1024 $\times$ 768 micro bolometer array\\
Pixel size	& 	 17 $\mu$m\\
Pixel scale at 20 km  &  10 m \\
Mass, total	& 	4 kg\\
Dimension	&  	180 $\times$ 150 $\times$ 120 mm \\
Volume	&  	283 cm$^3$\\
Operating temperature 	 &	-10$^\circ$C optimal\\
Temperature range		& -40 to +60$^\circ$C\\
Temperature stability	& $<$0.3$^\circ$C per minute\\
Total average power	&	10 W\\
Peak power	&	12 W\\
Pointing requirements	 &	1 mrad\\
\hline
\end{tabular}
\label{tab:TMC}
\end{table}

The heritage instrument, CMS, has dimensions of 380 $\times$ 335 $\times$ 185 mm, but can easily be reduced to 180 $\times$ 150 $\times$ 120 mm (Figure \ref{fig:TMC}, top).  The TMC radiometer approach has the advantage of a well-calibrated, straightforward data product, and significant flight heritage for high TRL.

\subsection{MBC body interior package}

This package probes the nucleus interior, and is made up of three instruments (two radars and the radio science investigation). The latter has no hardware component, but uses the spacecraft communication system to measure the MBC mass and internal mass distribution via Doppler ranging. The radar package consists of two radar units: SOURCE, a low frequency radar to probe the deep interior, and SSR, a higher frequency radar to fathom the near subsurface. Each radar system corresponds to different frequency ranges (15 -- 25 MHz and 300 MHz -- 3 GHz), and therefore to separated antennae and electronics. A joint ground segment for the two radars will result in a common interface for operations and for data handling. The ground segments inherit a large experience around Mars and 67P/Churyumov-Gerasimenko. 

\subsubsection{SOURCE: SOUnding Radar for Comet Exploration} 
The interior structure of a small body reflects its formation and evolution: Unlike planets, the self gravity of comets and small asteroids is not enough to compress and differentiate the rocky interiors. A wide range of interior structure can thus be expected depending on the applicable formation scenario, i.e. voids and macroscopic inhomogeneity as well as global homogeneity down to metre-scale. MBCs are particularly interesting targets for a study of interior structure, as their activity means that there has to be ice present, which has to be (mostly) buried to survive.

Low-frequency radio waves can penetrate both ice and soil (especially dry soil, with its low conductivity). This capability is exploited by ground penetrating radars, used for terrestrial applications such as location of buried pipes. Much of the transmitted radar energy is reflected back at the surface, but some propagates into the terrain with an attenuation dependent on the wavelength and material's dielectric and magnetic characteristics. Further reflections occur at interfaces between materials with different dielectric constants, thus allowing low-frequency radars to identify different geological layers. Information about the properties of the different layers can be inferred from: velocity of propagation in the medium (inversely proportional to the square root of the medium's dielectric permittivity), the attenuation within the medium and the fraction of energy scattered at the interface between two media.

SOURCE's design essentially matches SHARAD on {\it Mars Reconnaissance Orbiter} \citep{Croci2011} although the different operational environment will require some specialisation and simplification of electronics. It has two main physical elements: the electronics box and antenna. The former includes all transceiver electronics, signal processing and control functions. It comprises two separate electronic assemblies, The Transmitter and Front End (TFE) and Receiver and Digital Section (RDS), mounted on a support structure that acts as a radiator for thermal control and includes heaters and temperature sensors. The RDS is, in turn, divided into (1) the Digital Electronic Section, which carries out instrument control, communication with the spacecraft, timing generation, post-processing of the received data, and generation of the transmitted chirp signal; and (2) the receiver, which amplifies, filters and digitises the received signal. The TFE provides amplification of the transmitted signal, transmitter/receiver duplexing, and includes the matching network to interface with the antenna.

The self-standing TFE amplifies the low level chirp signals coming from the Digital Electronic unit and couples them to the dipole antenna; the unit also provides  the time division duplexing function, i.e., sharing the antenna between the transmitter and the receiver path. The SOURCE antenna is a 10-m dipole made of two 5-m foldable tubes, which when stowed are secured by a system of cradles and, when released, self-deploy because of their elastic properties. The main instrument parameters are reported in Table \ref{tab:SOURCE}.

\begin{table}
\caption{SOURCE parameters}
\begin{tabular}{lc}
\hline
Parameter & Value \\
\hline
Frequency band                   & 15--25 MHz chirp \\
Transmitter power                & 10 W \\
Pulse length                     & 85 $\mu$s \\
Receiving window length          & 135 $\mu$s \\
A/D converter sampling frequency & 26.67 MHz \\
Number of samples per echo       & 3600 \\
Pulse repetition frequency  & 700 Hz \\
Antenna (dipole)                 & 10-m tip-to-tip \\
Mass                             & 20.2 kg \\
Power                            & 34.7 W \\
\hline
\end{tabular}
\label{tab:SOURCE}
\end{table}

Both penetration and vertical resolution depend on the target's dielectric properties, while horizontal resolution and data rate depend on spacecraft altitude and relative velocity to the ground. Vertical resolution is inversely proportional to the square root of the relative dielectric permittivity of the comet material, which for a regolith, a porous rock or an ice/rock mixture should be  in the range 4--6. Given the SOURCE pulse bandwidth of 10 MHz, this translates in a depth resolution of 6--7.5 m \citep{Croci2011}. Penetration is more difficult to assess given natural materials' great variability in attenuation properties. SHARAD has been able to penetrate $>$1.5 km in dusty ice \citep[e.g.][]{Phillips2011}, and a few hundred metres in lava flows \citep{Carter2009}; we expect that MBC material is more similar to the former.

\subsubsection{Shallow Subsurface Radar, SSR}
The stability condition of volatile materials in MBCs remains largely unknown: what is the structure and the depth of the regolith covering the ice structures? Is the activation process driven by impact? What is the re-partition of such exogenous materials? {\it Rosetta} highlighted the complexity of 67P near surfaces structures and their impact on comet activity processes. 

SSR is a radar to perform the tomography of the first metres of the body in order to characterise the 3D structure of the regolith: layers, horizontal variability, pit or large buried cavities, near surface ice patches. This instrument will support the modelling of the activity scenarios and will contextualise remote sensing of surfaces in terms of structure and also for composition, as done with CONSERT on {\it Rosetta} \citep{Alain_permittivity}. 

SSR is a radar optimised for observations of small bodies at low operation altitude and low relative speed (few m/s). This configuration is less demanding than classical radar observations for planets at low or higher frequencies and allows an alternative design with limited resources. 
Its step frequency design offers a large versatility in terms of radar modes, resolution and operation distance. The nominal mode provides 2D images of the first 10 -- 50 m of the subsurface with only a few observation orbits. Combining optical surface and radar measurements of the shallow subsurface enables to discover changes in the material composition of the shallow subsurface with $\sim$metre horizontal resolution \citep{Herique2017}.

\begin{figure}
\begin{center}
\includegraphics*[width=\columnwidth]{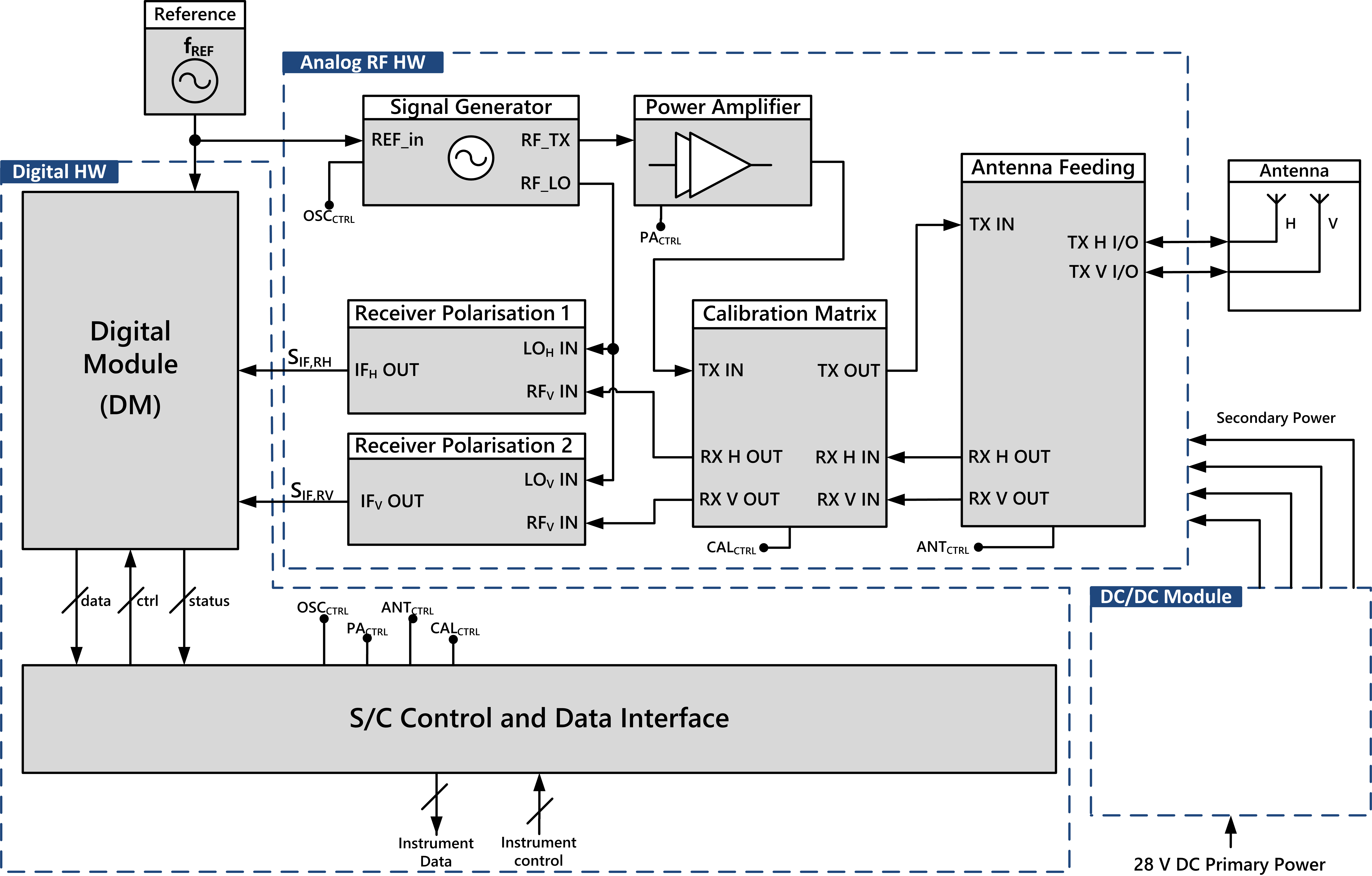}
\end{center}
\caption{SSR instrument block diagram.}
\label{fig:SSR-block}
\end{figure}

\begin{figure}
\begin{center}
\includegraphics*[width=0.3\columnwidth]{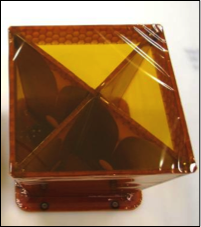}
\includegraphics*[width=0.3\columnwidth]{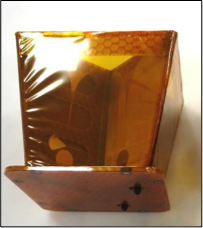}
\includegraphics*[width=0.3\columnwidth]{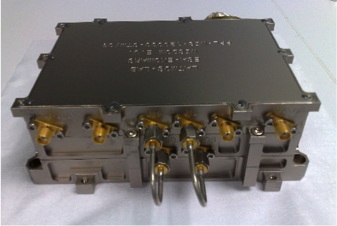}
\end{center}
\caption{SSR antenna and electronics box.}
\label{fig:SSR}
\end{figure}

Radar tomography is sensitive to observation geometries. The science objectives will be achieved with few arcs of orbits at low altitude ($<<$100 km) and low speed relative to the surface: the motion  comes mainly from MBC's own rotation. Within each arc, the radar is operating during a few hours in nadir direction, where the antenna is facing the body \citep{Herique2017}.

The SSR consists of an electronic unit and an antenna facing the surface (Figure \ref{fig:SSR-block} and \ref{fig:SSR}). The electronic unit delivers a set of adjustable frequencies from 300 MHz to 800 MHz in nominal mode and up to 3 GHz in extended mode. The synthesised signal is amplified and transmitted by the antenna system in circular polarisation. After the transmission of each single frequency, the echo is received by the perpendicular oriented antennas within a well-defined time window. Both perpendicular polarised signals are amplified, down-converted, sampled, phase-calibrated and combined to circular polarisation in a two-channel receiver chain. A micro controller inside a field-programmable gate array (FPGA) schedules radar operations, controls instrument settings and processing. On-board pre-processing is performed by a separate core inside the FPGA.
The antennas are to be accommodated on the spacecraft instrument deck. The frequency dependent half-power beam-width of more than 90$^\circ$ (in nominal mode) is limiting the pointing requirement to about $\pm 10^\circ$ or less. There are no other specific constraints as to thermal environment. 

\begin{table}
\caption{SSR resources and parameters}
\begin{tabular}{ll}
\hline
Antenna mass & 830 g\\
Total mass  & 2390 g\\
Power  max / mean & 137 W / 90 W \\
Data & 50 Gbit \\
Heritage & TRL 5-6\\
Modulation & Step Frequency\\
Carrier & 300 -- 800 MHz =$>$ 3 GHz\\
Resolution  & 1 m (3D)\\
Transmitter power & 20 W\\
Polarisation & Transmitter: 1 Circular;  \\
& Receiver: OC and SC\\
Noise Equivalent $\sigma_0$ & -40 dB.m2/m2\\
\hline
\end{tabular}
\label{tab:SSRparameters}
\end{table}

The radar inherits from the WISDOM  developed for the {\it ExoMars} rover mission (Phase B2 -- TRL 6 in 2016). It is also a carbon copy of the High-Frequency Radar instrument developed   within the scope of ESA {\it AIM} \citep{Patrick_Michel_AIM} phase A/B1\footnote{ESA contract RFP IPL-PTE/FE/yc/209.2015}. Ground segments for operation and for sciences will be based on the {\it Rosetta} CONSERT ground segment and on the Synthetic Aperture Radar processor developed in the frame of the  {\it AIM} phase A/B1.

For altitudes from 40 km down to 5 km the SSR works in synthetic aperture radar mode, mapping the surface and sub-surface backscattering coefficient. With many acquisitions from different geometries, this allows a 3D tomography with metre resolution, for a baseline frequency between 300 and 800 MHz. This Doppler resolution mainly comes from the rotation of the MBC \citep{Herique2017}. The penetration depth is expected to be 10 -- 50 m depending of ice/dust contents and porosity.
If the spacecraft approaches closer to the MBC than 5 km (not part of the baseline mission), at very low altitudes $<$500m, the reduced target distance would allow the choice of a set of radar parameters that could provide up to few tens of centimetres spatial resolution, while preserving a few metres penetration into the body, providing information about the body's internal shallow structure.
SSR also has the benefit of acting as a spacecraft altimeter with 6 cm resolution using the full bandwidth of 300MHz -- 3 GHz. This mode could support radio science mass determination and operations with on-ground processing, or possibly on-board calculations of the spacecraft surface distance (inverse Fourier transform), providing independent measurements in support of the attitude and orbit control system in automated hovering mode (see section \ref{sec:spacecraft}).
In jet mode, the SSR operates in an incoherent mode and would allow an estimation of the size/distribution of jet particles (size $>$mm). The performance/sensitivity of the incoherent mode is strongly dependent on jet size and density.

\subsection{MBC material and composition package}
The material and composition package directly samples the gas and dust coma of the comet, using in situ instrumentation. These instruments will primarily operate in the lowest altitude modes, as the low activity of MBCs dictates a close approach to have sufficient coma density. There are two instruments, one sensitive to gas and one to dust. The latter combines the measurement techniques used by GIADA and COSIMA on {\it Rosetta}, i.e. counts dust grains and analyses them, while the gas instrument is based on {\it Rosetta}/ROSINA, with an additional gas capture device based on {\it Philae}/PTOLEMY technology to increase efficiency.

\subsubsection{CAMS: Castalia Mass Spectrometer}
The {\it Castalia} Mass Spectrometer (CAMS) is dedicated to the characterisation of the MBC's gaseous environment. The key sensor is a 60~cm time-of-flight mass spectrometer with a gridfree reflectron to measure the abundances of volatiles in the comet's coma. CAMS will be able to derive isotopic ratios of the major volatiles and measure the inventory of the organic molecules in the coma of the comet - both crucial to address the science questions discussed in section \ref{sec:req}. The CAMS objectives and requirements are listed in table \ref{tab:CAMSreq}.

\begin{table*}
\centering
\caption{CAMS objectives}\label{tab:CAMSreq}
\begin{tabular}{p{0.3\linewidth}p{0.35\linewidth}p{0.3\linewidth}}
\hline
Scientific objectives & Associated critical measurement & Measurement requirements\\
\hline
Elemental gas abundances & e.g. separate CO from N$_2$ & Mass resolution*: $>$2500\\
Volatiles' molecular composition & Measure \& separate heavy organic molecules & Mass range: 2-250~amu/e\\
Isotopic composition of volatiles & e.g. separate $^{12}$CH and $^{13}$C. & \pbox{20cm}{Mass resolution*: $>$3000 \\ Relative accuracy $<$3\%} \\
Development of cometary activity & Measure the composition at different locations along the orbit, joint with CADS. & \pbox{20cm}{Mass range: 2-250~amu/e \\ Dynamic range per spectrum: 10$^{6}$ \\ Sensitivity: $>$10$^{-6}$~A/mbar} \\
MBC compositional heterogeneity & Mapping of active and inactive regions & Time resolution: 100~s\\
Coma chemistry + link surface and innermost coma & Measure neutrals and ions in the mass range of 2-250~amu/e over wide range of pressures. & \pbox{20cm}{Mass range: 2-250~amu/e \\ Total dynamic range: 10$^{8}$}\\
 \hline
\end{tabular}
\newline *at 50\% peak height 
\end{table*}

CAMS uses a storage ion source based on the {\it Rosetta} ROSINA-RTOF instrument \citep{RN13}. It has two redundant filaments with variable electron emission currents, depending on gas density. The ion source can be baked out to remove contaminants and evaporate semi-volatiles from grains hitting the ion source when close to the nucleus. The ion optical system (lenses and reflectron) uses ceramic and titanium composites and thus permits the lightweight and stable construction needed for a space mission. There are no internal grids in the reflectron, which allows high transmission and thus, together with the storage source, a very high sensitivity. CAMS uses fast detectors with an internal time response of $<$600~ps to detect single ions arriving within ns. The detector linearly amplifies the signal from incoming particles over a wide dynamic range. The ion source is enclosed in a combined structure which contains a flag function to determine background, and a protective dust cover. The flag is spring preloaded, contains mechanical end stops and switches and can be operated in three positions: open, closed, and sealed.

Based on {\it Rosetta} cruise phase experience, spacecraft outgassing is expected to interfere with scientific measurements at large heliocentric and MBC distances or at particularly low activity levels \citep{RN384}. Thus contamination by outgassing is addressed in all development phases and verified during cruise. The flag design takes into account that mass spectrometers are sensitive to decomposing lubricants. Lubricants are therefore excluded in the flag mechanism and mechanical wear is addressed by using appropriate materials (e.g. ceramics) for friction-loaded structures. The support structure of CAMS serves as the mechanical and thermal interface to the spacecraft and includes all CAMS electronics.

Also part of the instrument is a density sensor, CADS ({\it Castalia} Density Sensor) which, in addition to collecting scientific data, will autonomously turn on/off CAMS in case the operational pressure range of the mass spectrometer is exceeded. Its sensor head will be mounted such that it is in the free gas flow emanating from the MBC in order to minimise ram pressure influence. CADS is based on the ROSINA-Cometary Pressure Sensor (COPS): free electrons are emitted from a hot filament and accelerated toward the ionisation zone. Ionised atoms and molecules are collected by the cathode and detected by a sensitive electrometer to measure densities from 10$^{11}$-10$^4$~cm$^{-3}$. The measured ion current is directly proportional to the particle density in the ionisation volume of CADS. For redundancy CAMS and CADS are equipped with two filaments each.

\begin{figure}
\begin{center}
\includegraphics*[angle=-90, width=\columnwidth]{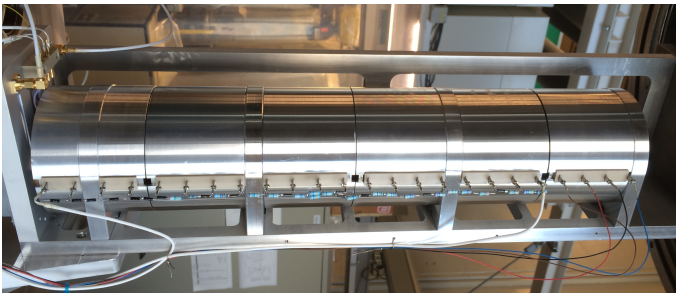}
\end{center}
\caption{CAMS prototype with drift section/reflectron.}
\label{fig:CAMS-sensor}
\end{figure}

Figure \ref{fig:CAMS-sensor} shows the CAMS laboratory prototype sensor with the drift section/reflectron. Ion source and detectors are located on the left side.

CAMS electronics will contain an instrument internal processor and all software, including optimisation and compression algorithms, will reside therein. Reduction of the produced data volume is therefore possible at hardware level (longer integration time for the mass spectrometer) and at software level (spectrum windowing, averaging, and compression). 

The CAMS calibration strategy is identical to that for ROSINA: A spare instrument serves as a ground reference model. The CASYMIR (CAlibration SYstem for the Mass Spectrometer Instrument ROSINA) facility is an ultra-high vacuum chamber with a molecular beam system \citep{RN611}. Water vapour beams mixed with organics are used to calibrate CAMS and CADS. The calibration of the mass spectrometer  for cometary ions is performed at the CASYMS (CAlibration SYstem for Mass Spectrometers) facility using a Low Energy Ion Source \citep{RN235}, where the mass spectrometer  is exposed to an ion beam of specified energy, energy spread, and composition.

Essential CAMS components are TRL 7 or higher and have heritage from a variety of projects: CAMS has direct heritage from the {\it Rosetta}/ROSINA RTOF where the detector design and the reflectron have been further developed. The electronics, controlling and data acquisition of the mass spectrometer has direct heritage from Luna Resurs and Lunar Glob. Also developed and tested to qualification levels are a protective dust cover combined with a flag to distinguish between spacecraft and comet outgassing. CADS is TRL 9 and is identical to the {\it Rosetta}/ROSINA COPS nude gauge.

Although CAMS can achieve the primary mass spectroscopy science goals of the mission, there is potential to develop an extra component, COUCH ({\it Castalia} Open University Concentrator Hardware), to increase its efficiency via collection and storage of volatiles for later analysis.  This unit exposes a cold gas collector material during the close operation phase of the mission for a few hours.  The collector doors are then closed to store the gas for analysis once the spacecraft has returned to parking orbit (possibly in parallel with communications windows, as no pointing would be required). Heating the collector material releases the adsorbed gas for analysis. Direct analysis by CAMS will in effect measure a concentrated gas sample collected over a long time period. Chemical processing of the gas sample will prepare for specific isotopic analysis such as D/H, $^{13}$C/$^{12}$C, $^{15}$N/$^{14}$N, and $^{18}$O/$^{17}$O/$^{16}$O.  Calibration gases are used to improve the isotopic accuracy and determine the sensitivity. Operation of the collectors during cruise and inactive phases of the mission will also determine the level of any spacecraft contamination. The use of chemical processing and calibration gases is based on technology from Ptolemy on the {\it Rosetta Philae} lander, which itself has a TRL of 8, but would require extra development to work on an orbiter and to feed the CAMS mass spectrometer (a larger format collector and doors). Such a system effectively increases the S/N ratio of the more challenging isotopic measurements without increasing the duration of the close distance operations of the mission. Software control of the package can be either through CAMS or independently with its own dedicated processor. On the CAMS side the gas inlet to the ion source would be based on the ROSINA Gas Calibration Unit that was successfully operated during the {\it Rosetta} mission \citep{RN13}.

\subsubsection{Combined dust impact detector and dust composition analyser: Dust Impact Detector and secondary Ion Mass Analyser (DIDIMA)}

\begin{figure}
\begin{center}
\includegraphics*[width=\columnwidth]{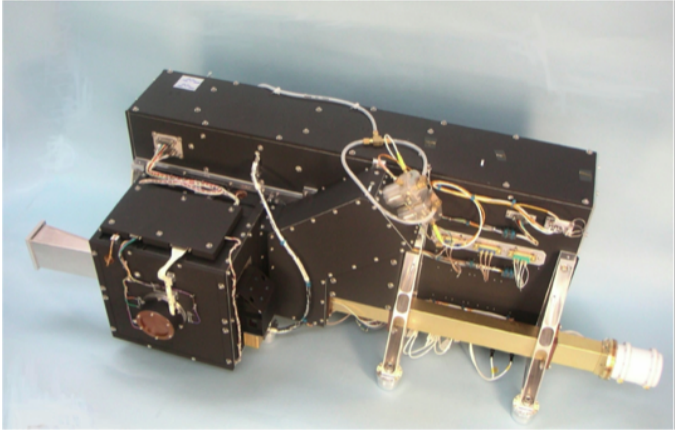}
\end{center}
\caption{Flight model of {\it Rosetta} COSIMA.}
\label{fig:COSIMA}
\end{figure}

\begin{figure}
\begin{center}
\includegraphics*[width=\columnwidth]{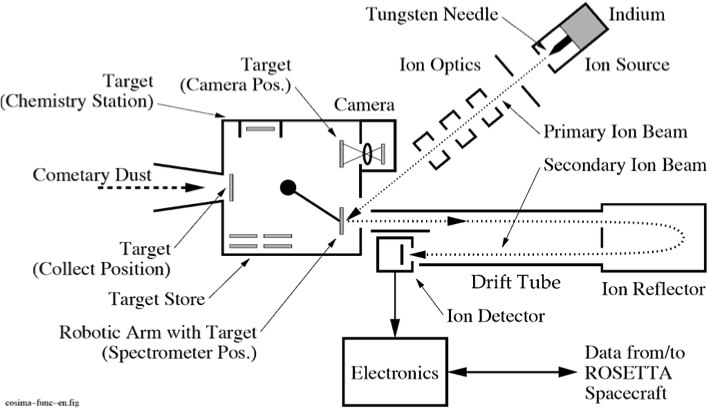}
\end{center}
\caption{Schematics of COSIMA operational subunits.}
\label{fig:COSIMA-schematic}
\end{figure}

The solid component of the MBC's coma, the dust, is made up of fragments of the solid surface and reflects the composition of rocky material in the nucleus. The chemical characterisation includes the main organic components, homologous and functional groups, as well as the mineralogical and petrographical classification of the inorganic phases. By studying the composition of MBCs' dust grains we can learn if they had a role in delivering to Earth the chemicals necessary for life as well as the properties of the earliest solids formed in the planet formation process. These types of measurement were performed at comet 67P by the COmetary Secondary Ion Mass Analyser (COSIMA) onboard {\it Rosetta}, the first instrument applying the secondary ion mass spectrometry in situ to cometary grains \citep{Kissel2007,Fray2016,Hilchenbach2016,Hornung2016,Langevin2016}. COSIMA (Figure \ref{fig:COSIMA}) consists of the following subunits: Primary Ion Source, Primary Ion Beam Subsystem, Time of Flight Spectrometer, Target Manipulator Unit, Ion Extraction, Ion Reflector, Ion Detector, Time to Digital Converter, Camera (Cosiscope), Control Electronics, Beam switch, Motherboard, Low voltage converter unit, CPU and interfaces, Chassis, Harness (Figure \ref{fig:COSIMA-schematic}). COSIMA collected dust on metal targets, imaged it routinely with a microscopic camera and analysed its composition covering a mass range 1-1200 amu with a mass resolution of 1400 at mass 100 amu. A pulsed primary indium ion beam gun shoots on the selected grains and releases secondary ions from the dust grain surface. The secondary ions, either positive or negative, are accelerated by an electrical field, drift through field-free regions and an ion reflector and counted by a microsphere plate detector according to their time-of-flight. The resulting mass spectra reflect the composition of the grain surface composition. 

\begin{figure}
\begin{center}
\includegraphics*[width=\columnwidth]{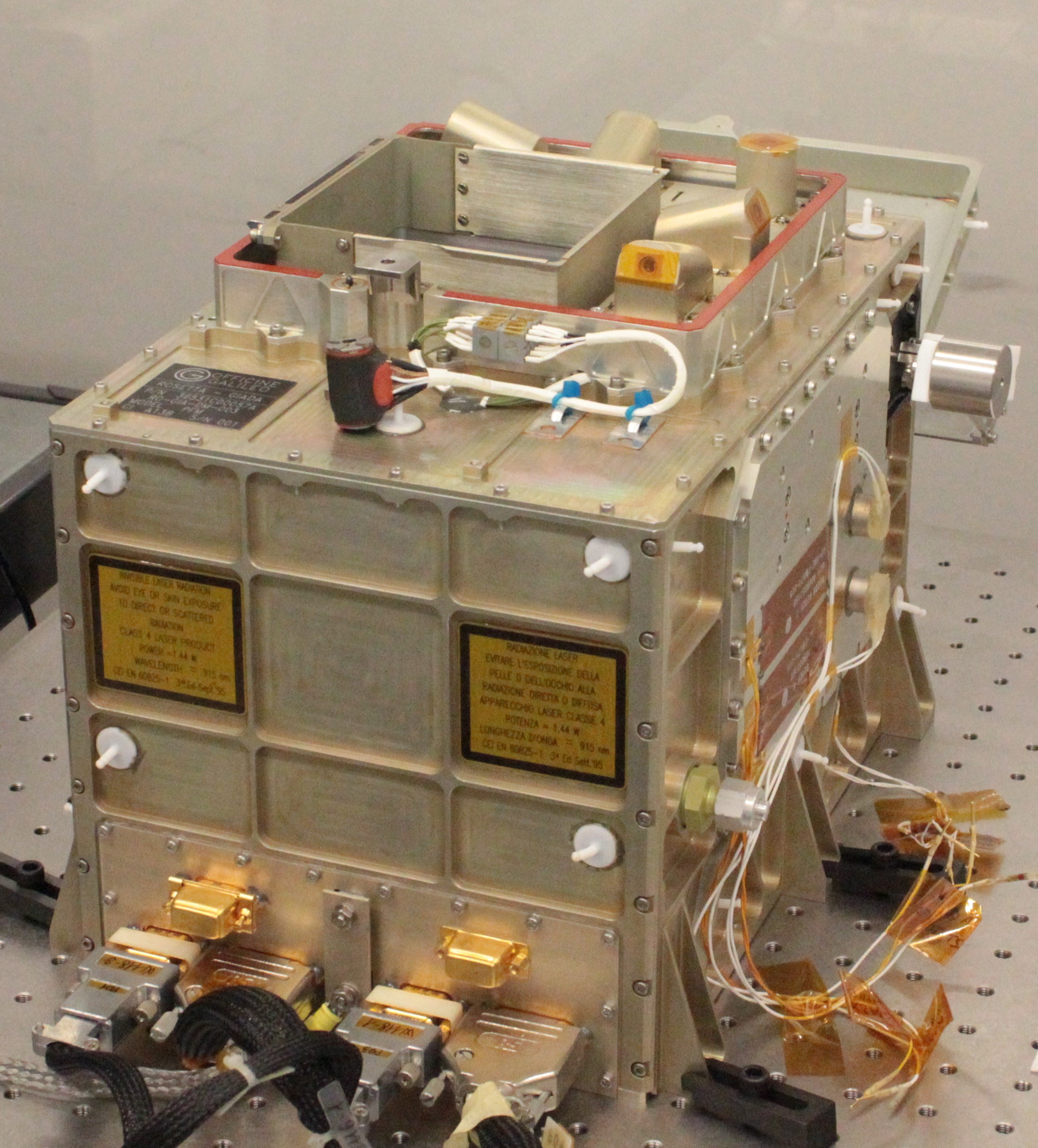}
\end{center}
\caption{The GIADA spare model housed in a clean room for calibration and testing
activities performed during the {\it Rosetta} cruise and scientific phase. On the top of the
instrument the squared baffle is the entrance of the dust particles that are detected first by the Grain Detection System (GDS), and then by the Impact Sensor (IS) mounted at the bottom of the instrument. The microbalances are the 5 cylinders distributed around the GIADA entrance.}
\label{fig:GIADA}
\end{figure}

\begin{table*}
\centering
\caption{GIADA measurement parameters}
\begin{tabular}{lccl}
\hline
Physical Quantity & \multicolumn{2}{c}{Measurement type} & 

Range\\
 & Direct & Derived &  \\
\hline
Speed & $\surd$ &  &   $1 - 300$ m s$^{-1}$\\
Momentum & $\surd$ &    & $6.5 \times 10^{-10} - 4 \times 10^{-4}$ kg m s$^{-1}$\\
Dust particle fluence & $\surd$ &    & $1.9 \times 10^{-9} - 2.9 \times 10^{-4}$ g cm$^{-2}$\\
Optical cross-section & $\surd$ &    & Function of particles' optical properties\\
Mass &  & $\surd$ &   $2.2 \times 10^{-12} - 4 \times 10^{-4}$ kg\\
Flux &  & $\surd$ &   $6 \times 10^{-12}$ g cm$^{-2}$ s$^{-1}$\\
Size &  &   $\surd$ & From cross-section \& grains' optical properties\\
 \hline
\end{tabular}
\label{tab:GIADA}
\end{table*}

\begin{figure}
\begin{center}
\includegraphics*[width=\columnwidth]{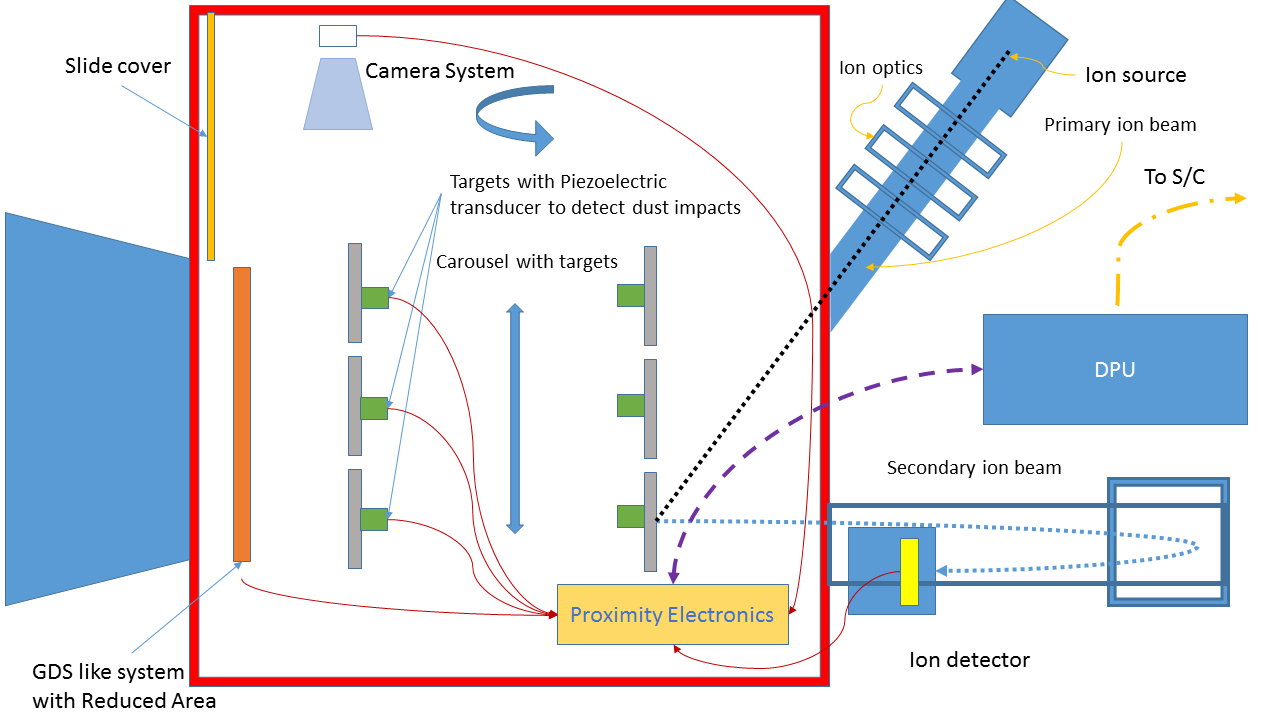}
\end{center}
\caption{Schematics of DIDIMA merged dust instrument}
\label{fig:DIDIMA-schematic}
\end{figure}

\begin{table}
\caption{GIADA and COSIMA versus DIDIMA resources}
\begin{tabular}{llll}
\hline
 & GIADA & COSIMA & DIDIMA\\
Mass [kg] & 7  & 20 & 23\\
Power [W] & 20 & 21 & 27\\
\hline
\end{tabular}
\label{tab:DIDIMA}
\end{table}

\begin{figure}
\begin{center}
\includegraphics*[width=\columnwidth]{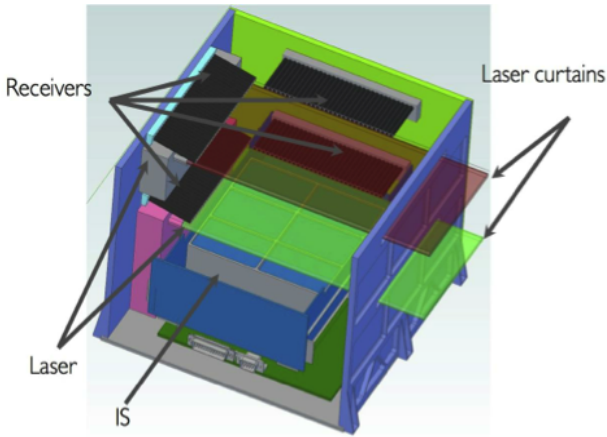}
\end{center}
\caption{Model of GIADA 2, the improved version of GIADA. In this updated instrument the GDS stage
consists of 2 laser curtains at different wavelengths (red and green boxes
indicated by the arrows). The light scattered by the particles entering the
instrument and crossing the 2 laser curtains allows  measurement of the particle speed and
cross section, providing also an indication on the particle optical properties. The
impact sensor placed below the GDS system measures the momentum of the  particle.}
\label{fig:GIADA2}
\end{figure}

The characterisation of the MBC dust environment includes the study of the dust physical and dynamical properties. The knowledge of momentum, speed, mass and density for individual dust particle as well as the global dust flux are necessary to depict the MBC coma dust environment. This aim was reached with the measurements performed by the Grain Impact Analyser and Dust Accumulator (GIADA), mounted on board {\it Rosetta} \citep{DellaCorte2014a,DellaCorte2016a,Colangeli2007}, in the coma of comet 67P \citep[e.g.][]{Rotundi2015,DellaCorte2015,DellaCorte2016b,Fulle2015,Fulle2016a, Fulle2016b}. GIADA (Figure \ref{fig:GIADA}) consists of three subsystems: 1) the Grain Detection System (GDS);  2) the Impact Sensor (IS); and 3) the Micro-Balances System. The GDS detects individual dust grains as they pass through a laser curtain, to measure their speed and  geometrical cross section, retrieved by means of calibration curves \citep{DellaCorte2016b} obtained using cometary dust analogues selected to get the extreme cases of dark and bright materials responses \citep{Ferrari2014}. The IS consists of a plate connected to five piezoelectric sensors measuring the dust grain momentum that, combined with the GDS detection time, provides a direct measurement of grain speed and mass. In addition, using GDS and IS combined measurements it is possible to constrain grains' bulk density. The Micro-Balances System is composed of five quartz crystal micro-balances providing the cumulative dust flux. A summary of the dust grain physical quantities that can be measured (or derived) by GIADA with the relative ranges of measurement is given in Table \ref{tab:GIADA}. These measurements can be predicted when in different dust environment by means of a simulation tool developed within the GIADA team \citep{DellaCorte2014b}. 

Taking advantage of the {\it Rosetta} heritage, a dust instrument for {\it Castalia} able to reach the scientific aims described above can be conceived combining the techniques adopted for COSIMA and GIADA. This would save resources and use  high TRL  components and methods. In the combined instrument, the Dust Impact Detector and secondary Ion Mass Analyser (DIDIMA), the COSIMA dust-collecting targets will be replaced by plates equipped with Piezoelectric sensors similar to the GIADA IS, to measure the dust particle momentum (Figure \ref{fig:DIDIMA-schematic}).  A grain detection system, similar to the GIADA GDS, at the DIDIMA entrance coupled with the IS of the collecting plate will allow derivation of the mass and speed of each individual particle entering the instrument. An upgrade of the GIADA subsystem is proposed, i.e. GIADA-2 (Figure \ref{fig:GIADA2}), which would combine two GDSs with two different laser wavelengths, set in cascade, and an IS with a higher momentum sensitivity with respect to the one currently used for GIADA. GIADA-2 would improve GIADA performances measuring particle size $>$ 0.5 $\mu$m, speed up to 500 m/s, momentum improved by a factor of 10. In addition, the position of the crossing particle through the laser curtain will be determined and combined with the position registered by the IS will give, rather than the speed, the velocity vector for each crossing particle, allowing the description of the dust grains velocity field. The two different laser wavelengths of the two GDSs plus the measurement of the scattered light from two different angles will give information on the dust optical constants thus constraining the grain composition analysed in parallel and more accurately by the secondary ion mass analyser of DIDIMA. DIDIMA will be able to perform the complete set of measurements performed by GIADA and by COSIMA separately during the {\it Rosetta} mission, achieving similar performance parameters but saving resources (see Table \ref{tab:DIDIMA}), and with the advantage that the velocity, mass and composition will be measured for the same grains. Although DIDIMA would have a lower TRL than re-flying direct copies of COSIMA and GIADA all its components have a high TRL, i.e. 5/6.

\subsection{MBC plasma environment package}
Similar to the {\it Rosetta} RPC consortium, the plasma package combines two separate small sensors. A common data processing unit and operational interface is foreseen.

\subsubsection{Charged particle Spectrometer for comets (ChaPS-C)}
The scientific aims of ChaPS-C are to monitor the solar wind at the MBC, detect the body's electric potential from the reflection and acceleration of charged particles at the surface, and to detect pickup ions originating at the MBC to establish the time of activity onset, and to then monitor activity levels. Similar measurements by the {\it Rosetta} RPC instruments found interesting physics during the first ever measurements of  low-level comet activity at the start of the mission \citep[e.g][]{Nilsson2015}; {\it Castalia} will probe even lower activity levels at a MBC. To achieve these aims, the instrument would detect electrons and both positive and negative ions from the solar wind and the MBC. Solar wind deflection and pickup ion fluxes will detect low and early MBC activity.

By measuring charged particle arrival directions, and energy per charge, ChaPS-C will determine the ion and electron density, temperature and velocity distribution functions of the MBC's local plasma environment. The baseline design will address this with a highly miniaturised electrostatic analyser combining both electron and ion detection, with a field of view (FOV) deflector system to allow electrostatic deflection of incoming ions and electrons by up to $\pm$ 45$^\circ$ out of the undeflected FOV plane.

\begin{figure}
\begin{center}
\includegraphics*[width=\columnwidth]{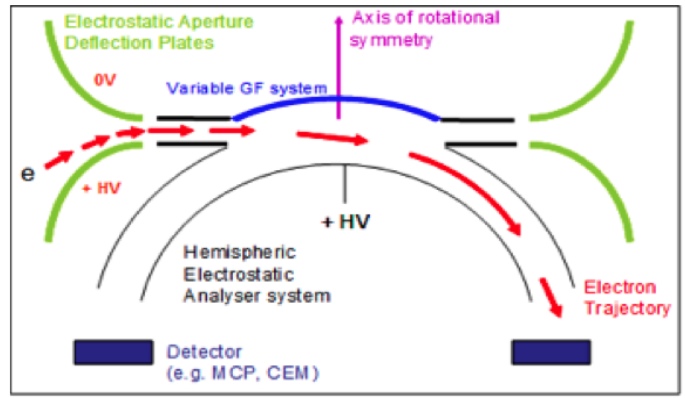}
\end{center}
\caption{Schematic of the enhanced performance top-hat design for SWA-EAS under development for Solar Orbiter.}
\label{fig:SWA-EAS}
\end{figure}

The sensor design draws on techniques and hardware developed for two major ongoing flight projects. The first, SWA-EAS, is an enhanced performance top-hat analyser 
for the ESA {\it Solar Orbiter} mission, shown schematically in Figure~\ref{fig:SWA-EAS}. Incoming charged particles enter the sensor through the exterior electrically grounded aperture grid. The particles are steered from the arrival direction into the Energy Analysis (EA) section using high voltages (HV) applied to either the upper or lower deflector electrodes providing a FOV Deflection System. The hemispherical EA section permits only charged particles of the selected energy and type, i.e., electrons or ions, to reach the detector subsystem. Charged particles transmitted through the EA are detected by a micro-channel plate (MCP) detector. The steering technique provides a considerably larger field of view, $\pm$20$^\circ$-60$^\circ$ x 360$^\circ$, compared to conventional top-hat designs which are typically 2$^\circ$-3$^\circ$ x 360$^\circ$, and provides an ideal solution for plasma monitoring on 3-axis stabilised spacecraft such as {\it Castalia}.

\begin{figure}
\begin{center}
\includegraphics*[width=\columnwidth]{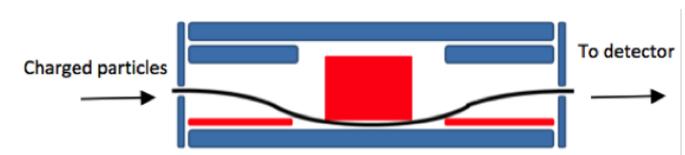}
\end{center}
\caption{ChaPS Bessel box geometry. Red are the electrodes that hold the voltage and blue are at ground.}
\label{fig:CHAPS-C-geom}
\end{figure}

An instrument developed more recently 
using an alternative geometry is the Charged Particle Spectrometer (ChaPS), developed and delivered for {\it TechDemoSat}, which returned data from low Earth orbit. The instrument consists of a number of miniaturised sensors using a modified Bessel box geometry, with a FOV of 1.5$^\circ$ x 20$^\circ$. The geometry is shown schematically in Figure~\ref{fig:CHAPS-C-geom}. The sensors are accommodated and optimised for different ion or electron plasma populations at Earth. 

\begin{figure}
\begin{center}
\includegraphics*[width=\columnwidth]{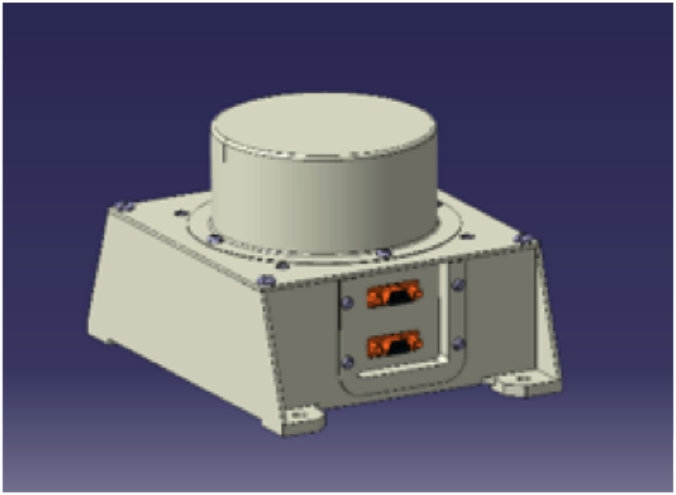}
\end{center}
\caption{Preliminary mechanical design of the ChaPS-C sensor.}
\label{fig:CHAPS-C-design}
\end{figure}

The proposed design for ChaPS-C will replace the top-hat section with the ChaPS sensor modified to provide the required FOV for {\it Castalia}, providing a low resource, compact analyser. A preliminary mechanical design of ChaPS-C consists of three modular elements (Figure~\ref{fig:CHAPS-C-design}), with the first two housed in the cylindrical section at the top and the third in the cubical section. The first module will consist of the charged particle optics elements, i.e., deflector plates and electrostatic analyser, and the second module the detectors (two back-to-back MCPs, each optimised for detection of  electrons or  ions) and front-end readout electronics. The cubical section will house other electronics: the high and low voltage power supplies, and the FPGA-based electronics board for sensor control, counters, data processing (compression and calculation of moments) and electrical interfaces to the spacecraft. A similar version of this design with a field of view of $\pm$22.5$^\circ$ x 120$^\circ$ has already been prototyped under an ESA contract. 

To cover electron energies that will include those accelerated due to the MBC negative surface potentials, as well as the potential observations of negative pickup ions such as O$^{-}$, the instrument's electron energy range will be 1~eV--3~keV. For positive ions, the energy range will be 5~eV--20~keV. This range will cover solar wind energies, as well as a significant energy range of newly-generated pickup ions resulting from the ionisation of outgassed MBC neutral species. The temporal resolution will be 2~s. The energy resolution $\Delta$E/E will be 16\% for electrons and 8\% for ions.

Plasma entering the analyser is deflected/energy analysed in the miniature analyser head and is detected by the MCP. The 16 segment readout anode is a modified version of the {\it Cluster} PEACE anode design whereby using 12 readout channels and using the unique coincidence technique developed on PEACE, 22.5$^\circ$ angular resolution is achieved for both the ion and electron channels. The analogue signals are fed to the FPGA where the coincidence logic and processing is applied and further processing to extract moments will be performed. The FPGA also provides the control for the sensor HV distribution and modulator stepping.
 
The current estimated mass of the instrument is 650~g, based on this preliminary design iteration and on experience with the ChaPS and SWA-EAS electronics, 
and 
its power consumption during operations is estimated at 1~W.
Although ChaPS is at TRL 8, a number of modifications are required, particularly to address specific field of view, accommodation and spacecraft issues, therefore ChaPS-C is currently at TRL 5/6. 
 
 \subsubsection{Magnetometer: MAG} 

\begin{figure}
\begin{center}
\includegraphics*[width=\columnwidth]{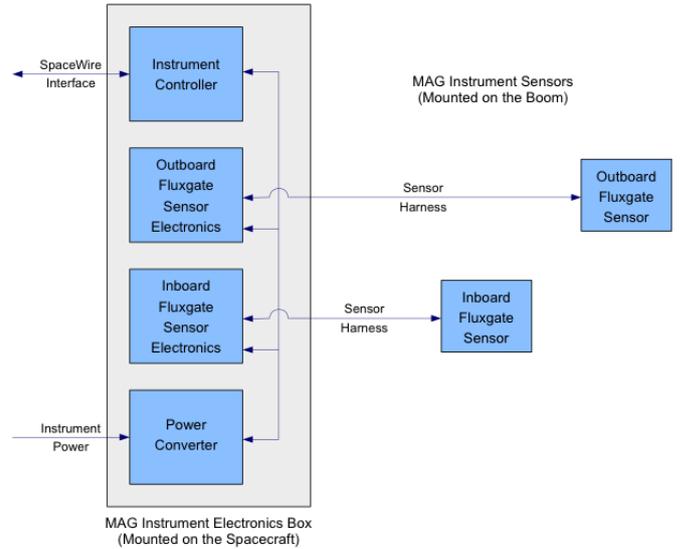}
\end{center}
\caption{MAG block diagram.}
\label{fig:MAG-block}
\end{figure}

The MAG instrument will measure the magnetic field in the vicinity of the MBC. This is crucial for establishing the magnetisation state of the body, searching for evidence of a dynamic atmosphere, and revealing how the comet interacts with the solar wind. 

Two separate digital fluxgate sensors will make magnetic field measurements. Fluxgate sensors are electrically passive and comprised of magnetically susceptible cores, each core is wrapped by two coils of wire. An alternating current is passed through one coil (the drive coil), cyclically driving the core to positive and negative magnetic saturation. A current proportional to the magnetic field along the coil axis is induced in the other coil (the sense coil). In addition, a current is applied through the sense coil to directly null the detected field along the coil axis through the magnetic core. The combination of current through the three sense coils orthogonal to each other thus allows the full, local magnetic field vector at each fluxgate sensor to be determined.

\begin{figure}
\begin{center}
\includegraphics*[width=\columnwidth]{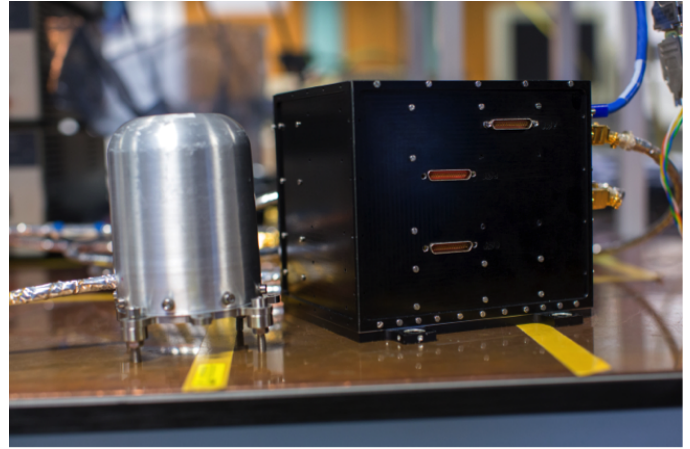}
\end{center}
\caption{{\it Solar Orbiter} MAG sensor head (left) and electronics box (right). Image: Imperial College London.}
\label{fig:MAG}
\end{figure}

The MAG instrument comprises two fluxgate sensors mounted on a spacecraft-provided rigid boom. Boom-mounting is required to reduce magnetic interference from the spacecraft. Two sensors are included for redundancy and calibration purposes, and lead to a relaxation of spacecraft electromagnetic cleanliness requirements at a fixed boom length. Instrument harnessing will attach both sensors to the spacecraft-mounted MAG electronics box, which contains a power converter and instrument controller and provides a link to the spacecraft bus via SpaceWire (Figs. \ref{fig:MAG-block} \& \ref{fig:MAG}). The instrument can be commanded simply with power-on/power-off and data rate selection commands. Calibration and science requirements lead to the need for operation of one or both sensors at Earth departure, continuously for at least 56 days prior to MBC orbit insertion, and continuously at the comet.

The range, resolution, and noise-level of this instrument are $\pm$64,000 nT, up to 8 pT (dependent on range), and $<$10 pT/Hz (at 1 Hz) respectively. The range is more than adequate to measure the geomagnetic field during an Earth flyby. Ranging of the instrument can be used to autonomously switch from measuring small magnetic fields at the limit of the fluxgate detectability, to large magnetic fields, whilst maintaining the same dynamic range of the instrument.

The mass of the sensors and electronics box is 4.5 kg. Each metre of harness will have a mass of $\sim$60 g. The electronics box volume is 30$\times$20$\times$20 cm and each sensor has a volume of 11$\times$11$\times$12 cm. The total MAG power requirement is 5.0 W, split between sensors and electronics. 
MAG can operate in normal and burst mode. Normal mode provides field measurements at 32 vectors/s and generates 2.44 kbits/s. Burst mode provides field measurements at 128 vectors/s and generates 9.21 kbits/s.

The outboard sensor is located at the extremity of the boom, furthest from the spacecraft. The inboard sensor is located closer to the spacecraft on the boom, of order 0.8 times the distance between the platform and the outboard sensor. The maximum tolerated DC spacecraft magnetic field at the outboard sensor is in the approximate range of 1-10 nT. MAG operating temperature is between -100$^\circ$C and 50$^\circ$C, and the sensors will survive down to -150$^\circ$C. The MAG sensors are not susceptible to radiation levels expected for this mission and the MAG electronics features a fully radiation hardened design. MAG requires a stable alignment between the sensor mounting, boom and boom mount on the spacecraft. The magnetometer has no demands on pointing of the spacecraft. 

Pre-flight on-ground calibration will be carried out at 
three axis coil and low temperature facilities, determining offsets, scale values, and orthogonality of each fluxgate sensor. Comparison between the sensors will be made. After launch, MAG should be the first instrument to be powered on during the deployment phase so that spacecraft fields generated by the rest of the science payload can be assessed. 
Operation during cruise is required to use Alfv{\'e}nic fluctuations in the interplanetary magnetic field to determine the time-varying offsets of each sensor.

All sub-units of the MAG instrument are TRL 8/9, and are space-qualified. Previous implementations of the fluxgate sensors and associated electronics have flown on missions such as {\it Cassini} and {\it Double Star}, and are included in the planned {\it Solar Orbiter} and {\it JUICE} magnetic field investigations, providing highly relevant, direct heritage \citep{Dougherty2004,Carr2005,OBrien2007}.

\subsection{Potential additional instruments}
The instruments above constitute a payload that would address {\it Castalia}'s primary science goals within the mission's design constraints. If sufficient mass and power are available, additional instruments would further enhance the already impressive scientific return. We list below additional experiments to be considered for inclusion during the mission study phase, resources allowing:
 \begin{itemize}
 \item Gamma ray and neutron spectrometer: Characterise MBC elemental composition and hydrogen content in the top surface layer (metre scale)  
\item Visible-Near-Infrared Imaging Spectrometer: Acquire reflectance and emission spectra of surface, dust, gas in visible and near-IR wavelength range 
\item Far UV imaging spectrometer: Acquire reflectance and emission spectra of surface, dust, gas in UV wavelength range
\item Sub-mm spectral imager/spectrometer: Acquire sub-mm radiation from MBC for thermal properties and water vapour detection.
\item Sub-surface science package (penetrator): In situ measurements of sub-surface physical conditions and possible chemistry. Miniaturised experiments in approximately increasing level of complexity would be: accelerometer, thermometer, magnetometer, seismometer, heat flux monitor, descent camera. With sample collection: sample imager, pH measurement, conductivity, chemical analysis, mass spectrometer. 
\item Surface package: Simple lander -- like {\it MASCOT} on {\it Hayabusa 2} \citep{Mascot} -- to measure surface physical properties (e.g. temperatures, thermal inertia, magnetic field, mineralogy, small scale structure). The mass of such a lander could be as low as 15 kg. {\it MASCOT-2} and even smaller CubeSat-based landers were also considered for {\it AIM}.
\item Polarimeter: Obtain additional information about surface properties (e.g. grain size) from polarisation of reflected light in visible wavelength range.
\item Volatile in situ Thermogravimetry Analyser: Characterise the dust/ice ratio of the coma grains and the presence of volatiles (water/organic) bound to the crystal structure of the refractory compounds \citep{VISTAref}.
\end{itemize}

\section{Castalia spacecraft}\label{sec:spacecraft}

The mission architecture and (sub)system design of {\it Castalia} was initially developed in 2013 during a Concurrent Engineering study at DLR in Bremen. An international team of scientists was combined with system and subsystem engineers, as well as cost experts from OHB System AG and DLR. The mission was proposed for the ESA M4 call in early 2015 \citep{M4abs1,M4abs2}, but it was not selected. For the M5 call the mission has been revised, reducing the system complexity and cost, but ensuring compliance to the overarching scientific objectives, mission programmatics and margin philosophy. The original mission architecture and multiple `hovering' manoeuvres were  retained. The spacecraft arrives before the start of the expected activity at the MBC. One year of nominal science operations is foreseen to perform all the scientific activities and return the associated data with a comfortable margin.

\begin{figure}
\begin{center}
\includegraphics*[width=0.8\columnwidth,trim=0 0 180 0,clip]{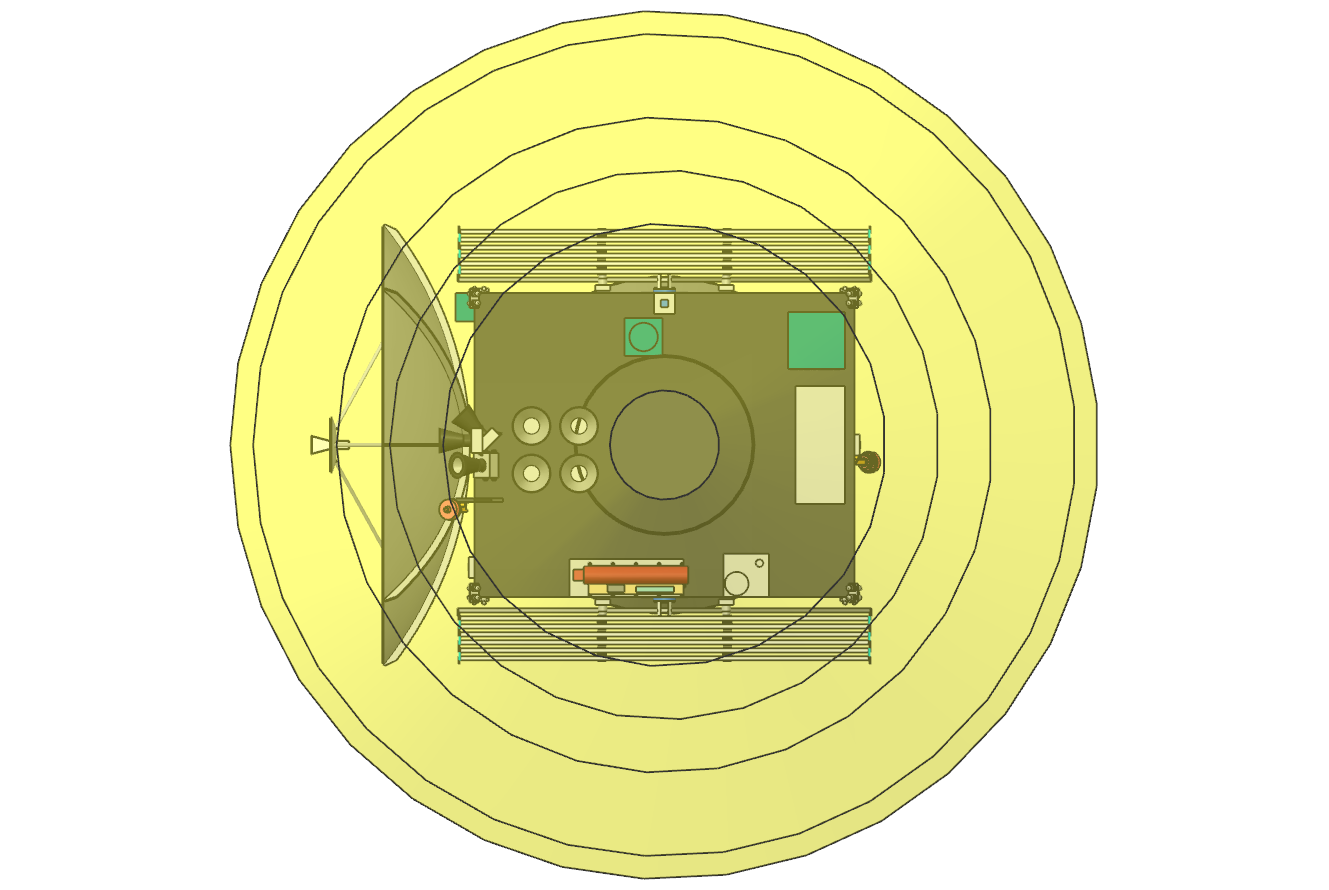}\\
\includegraphics*[width=0.18\columnwidth,trim=550 0 550 0,clip]{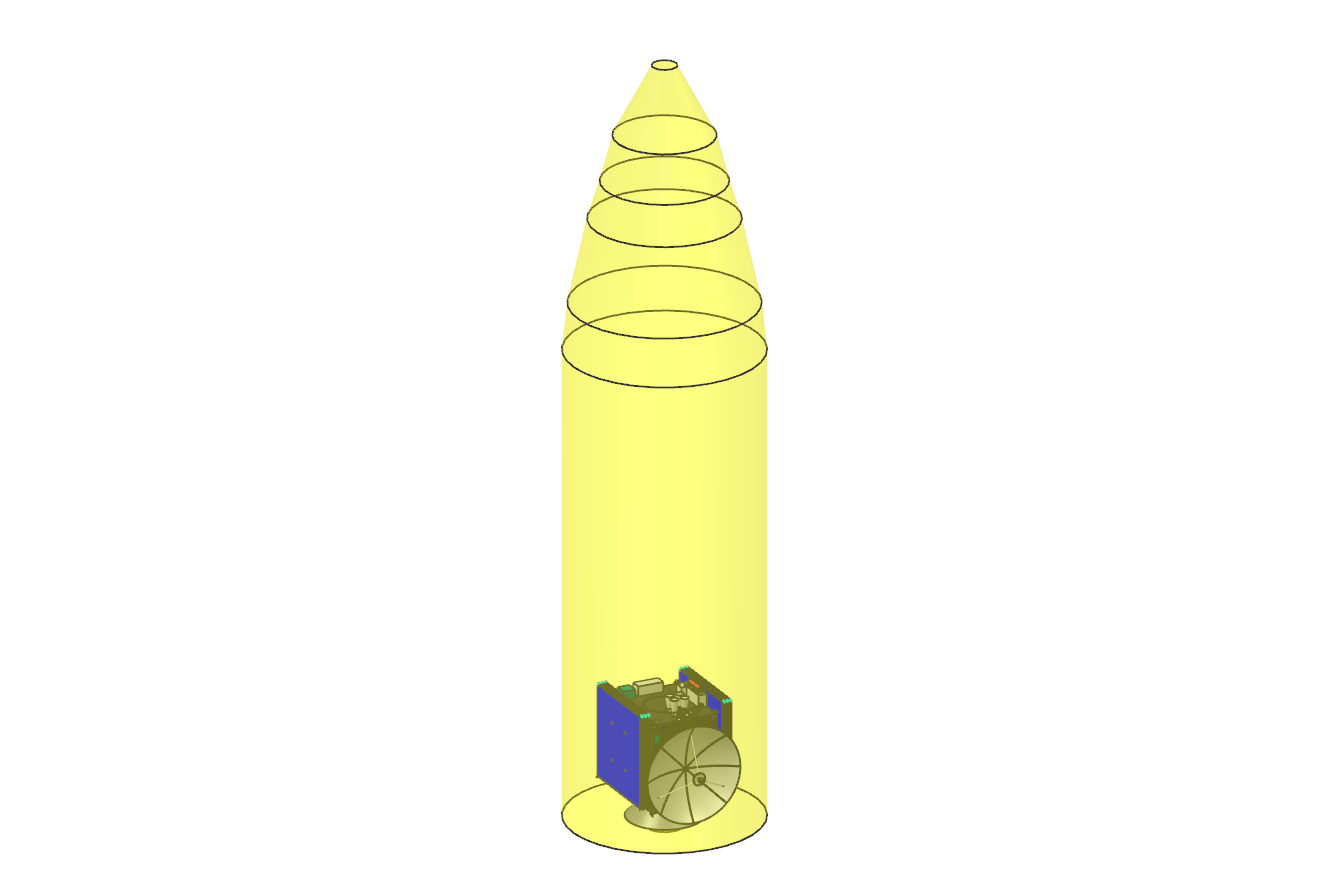}
\includegraphics*[width=0.8\columnwidth,trim=250 0 0 0,clip]{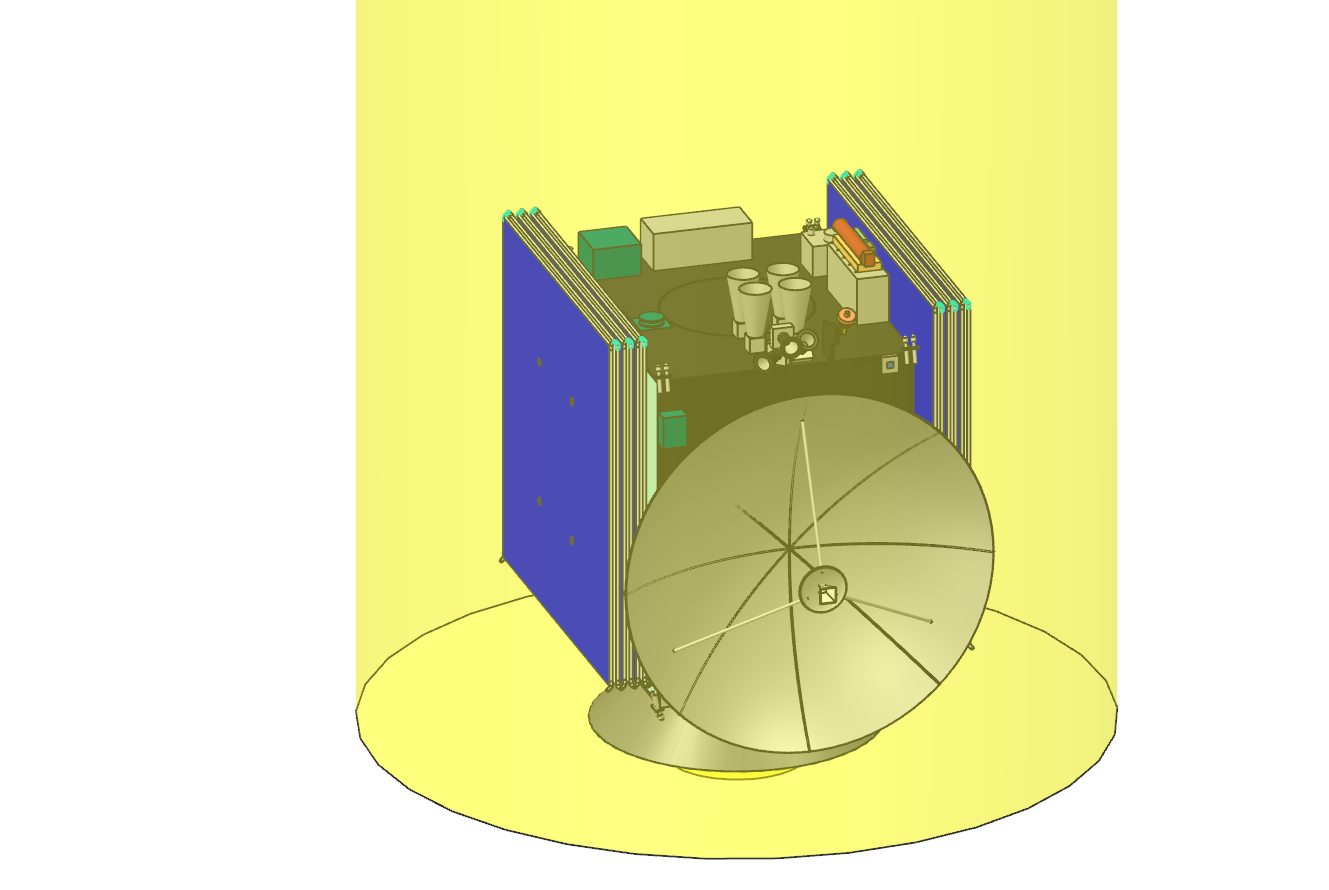}
\end{center}
\caption{Spacecraft in its Stowed Configuration under the Ariane 6.2 Fairing}
\label{fig:spacecraft-stowed}
\end{figure}

\begin{figure*}
\begin{center}
\includegraphics*[width=\textwidth,trim=0 200 0 200,clip]{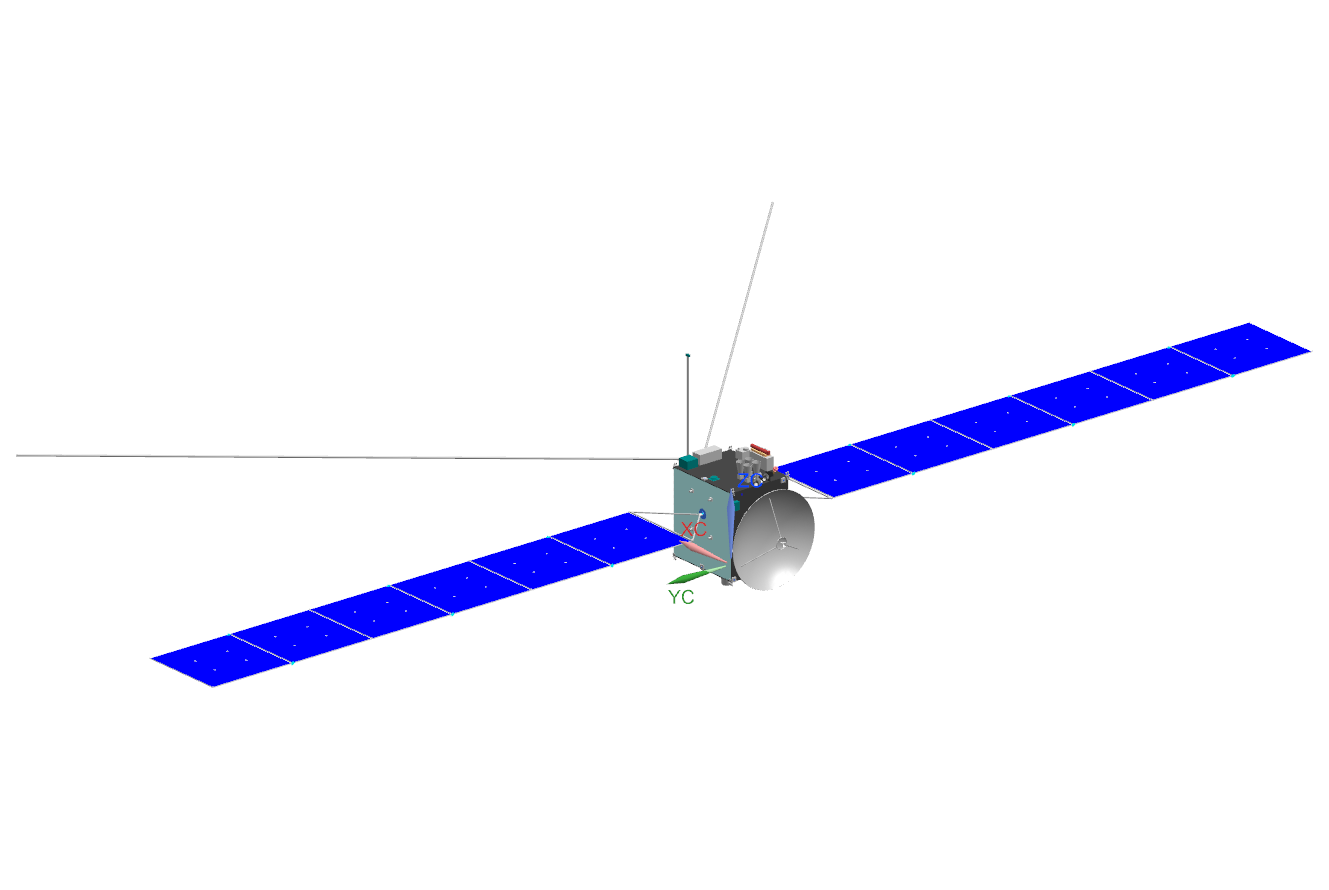}
\end{center}
\caption{Spacecraft in its Deployed Configuration}
\label{fig:spacecraft}
\end{figure*}

\begin{figure}
\begin{center}
\includegraphics*[width=\columnwidth]{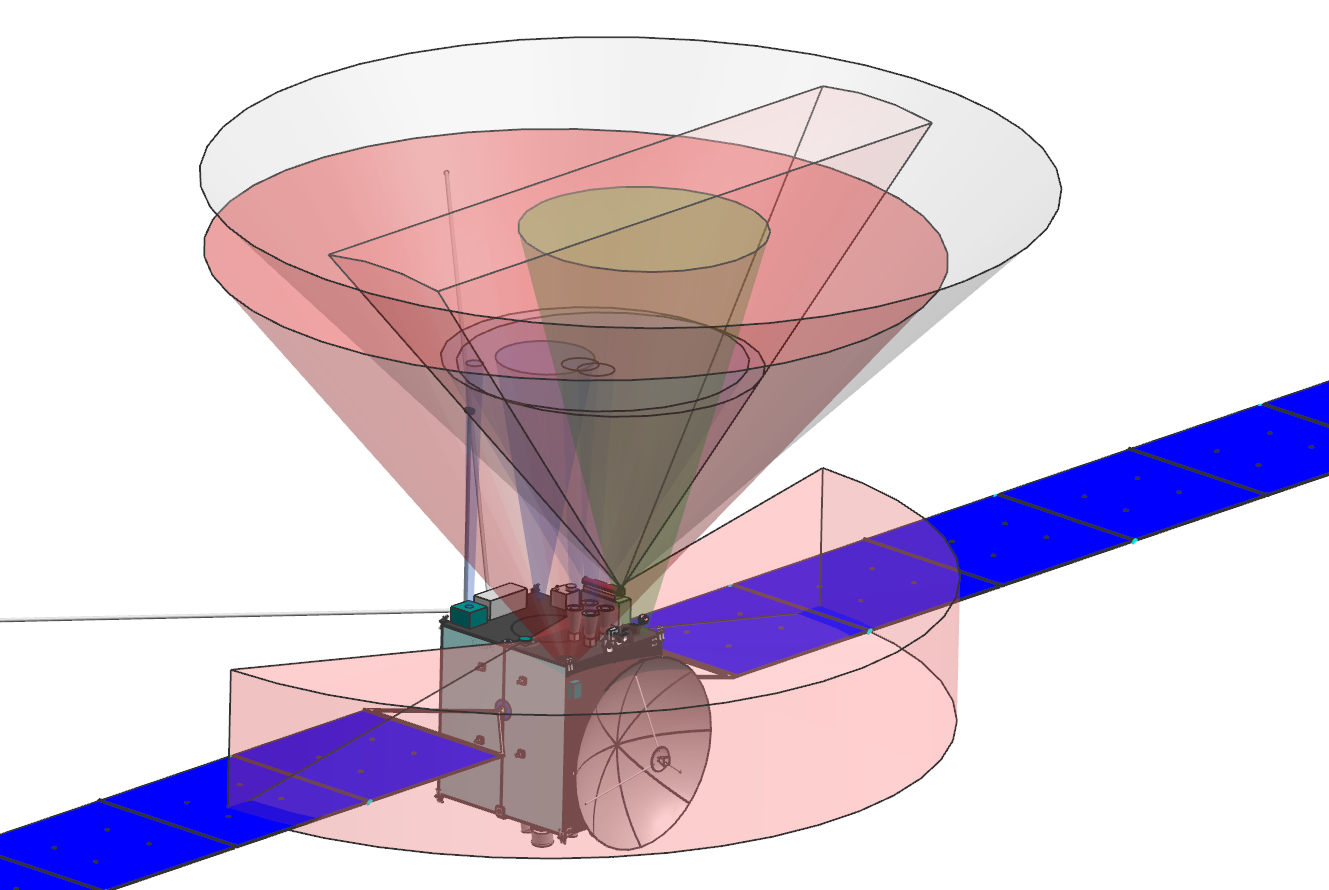}
\end{center}
\caption{{\it Castalia} spacecraft, with FOV of remote sensing instruments indicated.}
\label{fig:FOVs}
\end{figure}

Figure \ref{fig:spacecraft-stowed} shows the spacecraft in its launch configuration, under the Ariane 6.2 fairing. At the time of writing (and when preparing the M5 proposal) the performance of the Ariane 6.2 in providing a direct escape trajectory was not known. The mission has therefore been designed to be compliant with the uplift capability of the Soyuz launch vehicle from Kourou. This represents the worst-case conditions and limits the spacecraft's maximum wet mass at launch.  All other parameters were taken from the draft issues of the Ariane 6.2 user manual \citep{Ariane-6-manual}. Figures \ref{fig:spacecraft} and \ref{fig:FOVs} show the spacecraft in its deployed, in-flight configuration and the field-of-view of the scientific instruments. The spacecraft has a dry mass of about 1.2 tonnes including margins. An additional mass of about 30 kg of hydrazine is needed for the attitude control and hovering purposes, and 320 kg of xenon is need for the transfer trajectory, both including generous margins. The resulting launch mass is about 1.6 tonnes, including a standard launch adapter. For the M5 version various efficiencies saved about 115 kg compared to the M4 proposal.

The spacecraft is powered by two large deployable solar arrays. Their size is driven by the EP during the transfer when performing a manoeuvre. The power demand during the communication and science phases at the MBC are significantly lower. The main structure of the spacecraft consists of a load bearing central tube. This accommodates the propellant tanks (xenon and hydrazine) and is connected to the launch adapter at the bottom. Shear panels increase the stiffness of the spacecraft and provide separate compartments for internal thermal zones. Optical solar reflectors, radiators, heaters, multi-layer insulation, thermistors and doublers are used to control the thermal control of the spacecraft. The radiator and heater power were sized according to the equipment and instrument temperature requirements.  An electric propulsion system (RIT-2X) is used to perform the orbital transfer. A mono-chemical propulsion system (hydrazine thrusters) performs all operations close to the target. 

Navigating in the far and close proximity fields is achieved by Narrow and Wide Angle Cameras respectively. A Light Detection and Ranging (LIDAR) system is also used at close proximity to break the inherent range ambiguity of the optical sensors. A standard set of star trackers, Sun sensors and an inertial measurement unit provide attitude and attitude rate estimation. Reaction wheels provide attitude control. Attitude, orbit control and wheel desaturation is provided by hydrazine thrusters. Repetitive hovering for in situ scientific measurements is the most critical  phase. The acquisition of scientific data occurs without perturbation by thruster firing and after the spacecraft's settling period. Hovering occurs autonomously by the spacecraft and includes the capability for a retreat manoeuvre. Direct control from the Earth is neither necessary nor possible due to the dynamics of the activity and the distance to Earth.

The communication subsystem consist of an X-band system with a high gain antenna and high power traveling-wave tube amplifier. Two low gain antennas and a medium gain antenna provides communication during the early mission phases and emergency cases respectively. Weekly communication slots during the transfer and science operations remove the need for an antenna pointing mechanism. Data downlink occurs via the ESA ESTRACK ground stations. The high science data rate in the orbiting phase drives the onboard data handling system, where data is downlinked regularly. The size of the mass memory is limited to the storage of two cycles of science data.  

The proposed spacecraft design is based on existing European technology, with no international collaboration currently foreseen. The spacecraft will also benefit from ongoing technology developments. Ground control operations at comet 67P from {\it Rosetta}, for example, and studies such as PHOOTPRINT \citep{ESA2015a,ESA2015b}, {\it AIM}, {\it Marco-Polo-R} and {\it JUICE} can help simplify the automated real-time navigation operations at the MBC. Experience from {\it Rosetta} and ground-based laboratory experiments \citep{gibbings2014} have demonstrated the limiting effects of the sublimation products on the solar arrays and any exposed surface in general, during the close proximity operations. Furthermore, the expected level of sublimation by the MBC is orders of magnitude lower than experienced in the {\it Rosetta} mission at 67P. {\it Rosetta}, in fact, only experienced a 2 \% degradation in its solar cells. The degradation expected in a year at an MBC is therefore expected to be negligible. These benefits can also be combined with developments and integration of: 1) image-processing algorithms for providing autonomous, feature recognition, visual-based navigation cameras;  2) improved radar technology;  3) improved LIDAR altimeter for close proximity manoeuvres (e.g. {\it MarcoPolo-R}, Mars) and 4) improved closed-loop guidance, navigation and control algorithms with real-time validation and verification.

\section{Conclusions}\label{sec:conclusions}
We have described a mission to rendezvous with a MBC, with a suitable scientific payload to explore this new population, test theories on how their unexpected activity is driven, and directly sample the water in the asteroid belt. A detailed study with engineers from OHB System AG (Bremen) found that a suitable spacecraft could be built, launched and operated within the budget of an ESA medium class mission. This mission, {\it Castalia}, has been proposed to the M5 call of the ESA Cosmic Vision programme, with an expected launch in the late 2020s and operations at the MBC 133P in the 2030s. The instrument payload comprises four packages, made up of: visible/near-IR cameras, a thermal imager, deep and shallow sub-surface penetrating radars, neutral and ion mass spectrometers, a combined dust detector and analyser, a plasma package and a magnetometer. This mission will provide important constraints on Solar System formation and evolution theories by quantifying the amount of ice present today in a small body. It also has significant astrobiology implications, as it is possible that the water ice buried within MBCs samples the same original population that brought water to Earth.

\section*{Acknowledgments}
We thank DLR Bremen and OHB System AG for their hospitality and support during the concurrent design facility study that shaped the mission design.
CS is funded by the UK Science and Technology Facilities Council through a Ernest Rutherford fellowship.
Fig.~\ref{fig:133Pimage} is based on observations collected at the European Organisation for Astronomical Research in the Southern Hemisphere under ESO programme 184.C-1143.
We thank Gonzalo Tancredi and an anonymous referee for comments that improved this paper.

\section*{References}

\bibliographystyle{model2-names.bst}\biboptions{authoryear}
\bibliography{Castalia-ASR}

\end{document}